\documentclass[pre,twocolumn,amsmath,amssymb,showpacs,preprintnumbers]{revtex4-1}
\usepackage{mathrsfs}
\usepackage[dvipdfmx]{graphicx}
\usepackage{dcolumn}
\usepackage{bm}
\usepackage{color}

\newcommand{\ra}{\rangle}
\newcommand{\la}{\langle}
\newcommand{\tb}{\textbf}

\newcommand{\sgn}{\mathrm{sgn}}
\newcommand{\heta}{\hat{\eta}}

\newcommand{\mrR}{\mathrm{R}}
\newcommand{\mrT}{\mathrm{T}}

\newcommand{\CM}{{\mathrm{c.m.}}}
\newcommand{\Dp}{\Delta p}

\newcommand{\htau}{\tau}

\newcommand{\ve}{\varepsilon}

\newcommand{\hr}{r}
\newcommand{\tr}{\tilde{r}}

\newcommand{\pst}{{\mathrm{post}}}

\newcommand{\Dtc}{\Delta t_{\mathrm{can}}}
\newcommand{\rmid}{r_{\mathrm{mid}}}
\newcommand{\mutot}{\mu_{\mathrm{tot}}}

\begin{document}

\title{Derivation of the Boltzmann Equation for Financial Brownian Motion:\\ Direct Observation of the Collective Motion of High-Frequency Traders}

\author{Kiyoshi Kanazawa$^{1,2}$}\email{Corresponding author: kanazawa.k.ae@m.titech.ac.jp}
\author{Takumi Sueshige$^{2}$}
\author{Hideki Takayasu$^{1,3}$}
\author{Misako Takayasu$^{1,2}$}

\affiliation{
	$^1$Institute of Innovative Research, Tokyo Institute of Technology, 4259 Nagatsuta-cho, Midori-ku, Yokohama, 226-8502, Japan\\
	$^2$Department of Mathematical and Computing Science, School of Computing, Tokyo Institute of Technology, 4259 Nagatsuta-cho, Midori-ku, Yokohama, 226-8502, Japan\\
	$^3$Sony Computer Science Laboratories, 3-14-13 Higashi-Gotanda, Shinagawa-ku, Tokyo, 141-0022, Japan
}

\begin{abstract}
	A microscopic model is established for financial Brownian motion from the direct observation of the dynamics of high-frequency traders (HFTs) in a foreign exchange market. 
	Furthermore, a theoretical framework parallel to molecular kinetic theory is developed for the systematic description of the financial market from microscopic dynamics of HFTs. 
	We report first on a microscopic empirical law of traders' trend-following behavior by tracking the trajectories of all individuals,
	which quantifies the collective motion of HFTs but has not been captured in conventional order-book models.
	We next introduce the corresponding microscopic model of HFTs and present its theoretical solution paralleling molecular kinetic theory: 
	Boltzmann-like and Langevin-like equations are derived from the microscopic dynamics via the Bogoliubov-Born-Green-Kirkwood-Yvon hierarchy.
	Our model is the first microscopic model that has been directly validated through data analysis of the microscopic dynamics, exhibiting quantitative agreements with mesoscopic and macroscopic empirical results. 
\end{abstract}
	

\maketitle
	\begin{figure*}
		\centering
		\includegraphics[width = 180mm]{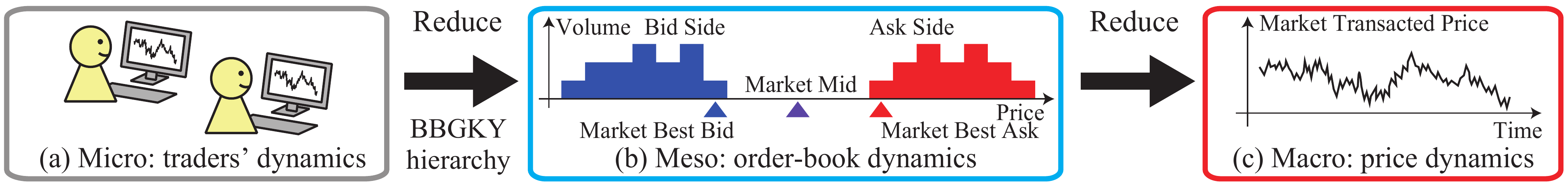}
		\caption{Hierarchies for financial Brownian motions for the microscopic, mesoscopic, and macroscopic dynamics.}
		\label{fig:hierarchy_fin}
	\end{figure*}
	{\it Introduction.--}
	In physics, the study of colloidal Brownian motion has a long history beginning with Einstein's famous work~\cite{Einstein1905};
	the understanding of its mechanism has been systematically developed in kinetic theory~\cite{Chapman1970,VanKampen}. 
	Specifically, from microscopic Newtonian dynamics, the Boltzmann and Langevin equations are derived for the mesoscopic and macroscopic dynamics, respectively. 
	This framework is a rigid foundation for various nonequilibrium systems (e.g., active matter, granular gas, Feynman ratchets, and traffic flow~\cite{Bertin2006,Brilliantov,Broech2004,Broeck2006,Helbing,Nishinari,Prigogine}), 
	and its direct experimental foundation has been revisited because of recent technological breakthroughs~\cite{Raizen2011,Raizen2010}. 

	In light of this success, it is natural to apply this framework beyond physics to social science~\cite{Pareschi2013}, such as finance. 
	Indeed, the concept of random walks was historically invented for price dynamics by Bachelier earlier than Einstein~\cite{Bachelier}, 
	and its similarities to physical Brownian motion (e.g., the fluctuation-dissipation relation) are intensively studied by recent high-frequency data analysis~\cite{Yura2014}. 
	As an idea in statistical physics, the dynamics of financial markets are expected to be clarified from first principles by extending kinetic theory. 

	Although this idea is attractive, the kinetic description has not been established for financial Brownian motion. 
	Why has not this idea been realized yet? 
	In our view, the biggest problem is the absence of established microscopic models; 
	there exist empirical validations of mesoscopic~\cite{Slanina2014,Maslov2000,Daniels2003,Smith2003,Bouchaud2002,Farmer2005,Yura2014} and macroscopic models~\cite{Mantegna1995,Mantegna2000,Plerou1999,Lux1996,Guillaume1997,Longin1996,PUCK}, 
	whereas no microscopic model has been validated by direct empirical analysis. 
	Indeed, previous microscopic models~\cite{Kyle1985,Takayasu1992,Bak1997,Lux1999,Yamada2009} were purely theoretical and have no quantitative evidence microscopically. 
	To overcome this crucial problem as an empirical science, two missing links have to be connected: (i)~establishment of the microscopic model by direct observation of traders' dynamics (Fig.~\ref{fig:hierarchy_fin}a) 
	and (ii)~construction of a kinetic theory to show its consistency with mesoscopic and macroscopic findings (i.e., the order-book and price dynamics (Fig.~\ref{fig:hierarchy_fin}b, c)). 

	In this Letter, we present the corresponding solutions by direct observation of high-frequency trader (HFT) dynamics in a foreign exchange (FX) market:
	(i)~a microscopic model of HFTs is established by direct microscopic evidence, and 
	(ii)~corresponding kinetic theory is developed to show its consistency with mesoscopic and macroscopic evidence. 
	We analyzed order-book data with anonymized trader identifiers (IDs) to track trajectories of all individuals.
	We found an empirical law concerning trend following among HFTs, which has not been captured by previous order-book models.  
	Remarkably, this property induces the collective motion of the order book and naturally leads the layered order-book structure~\cite{Yura2014}. 
	We then introduce a corresponding microscopic model of trend-following HFTs.
	Starting from their ``equations of motion,"  
	Boltzmann-like and Langevin-like equations are derived for the order-book and price dynamics. 
	A quantitative agreement is finally shown with our empirical all findings. 
	Our work opens the door to systematic descriptions of finance based on microscopic evidence. 
	\begin{figure*}
		\centering
		\includegraphics[width=180mm]{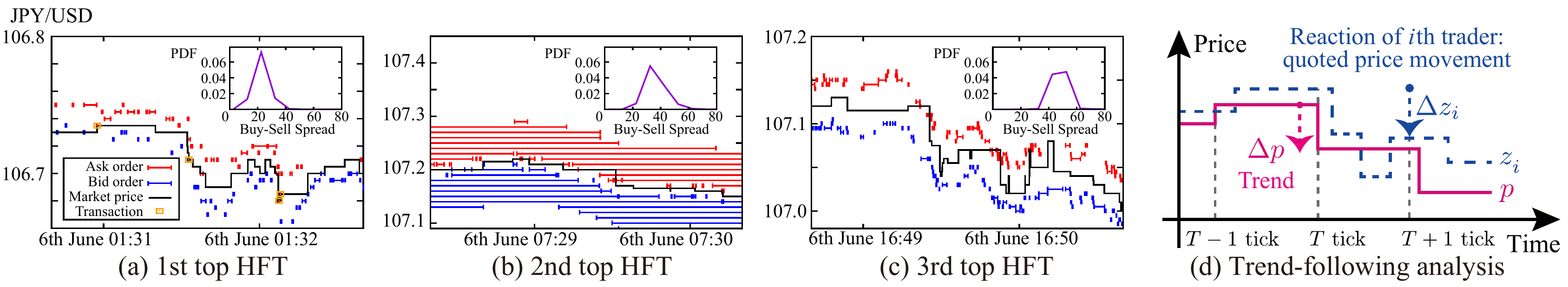}
		\caption{
					(a--c)~Lifetimes of orders are plotted as trajectories for the top 3 HFTs. 
					Typical traders tend towards continuous two-sided quotes,
					with the buy-sell spread fluctuating around a time constant unique to the trader. 
					The percentage of two-sided quotes among HFTs was 48.4\% (see Appendix~\ref{app:two-sided}). 
					(d)~Quantification of trend following for individual traders, 
					where $\Delta p$ and $\Delta z_i$ are the movements of the market price and the midprice of the $i$th trader, respectively. 
				}
		\label{fig:trajectory_traders}
	\end{figure*}
	
	{\it Observed microscopic dynamics.--}
	We analyzed the high-frequency FX data between the U.S. dollar (USD) and the Japanese Yen (JPY) on Electronic Broking Services for a week in June 2016 (see Appendix~\ref{sec:MarketRule}). 
	The currency unit used in this study is 0.001 yen, called the tenth pip (tpip). 
	Here we particularly focused on the dynamics of HFTs~\cite{Menkveld2011}, frequently submitting or canceling orders according to algorithms (see Appendix~\ref{app:def_HFT}). 
	The typical trajectories of bid and ask quoted prices are illustrated in Fig.~{\ref{fig:trajectory_traders}a--c} for the top 3 HFTs. 
	They modify their quoted prices by successive submission and cancellation at high speed typically within seconds; 
	almost 99\% of their submissions were finally canceled without transactions (see Appendix~\ref{app:cancel}). 
	With the two-sided quotes they also play the role of liquidity providers~\cite{Menkveld2013,EBSRule} according to the market rule, keeping the balance between the bid and ask order book. 
	Buy-sell spreads, the difference between the best bid and ask prices for a single HFT, were observed to fluctuate around certain time constants (see the insets for their distributions and Appendix~\ref{app:buy-sell}).

	\begin{figure*}
		\centering
		\includegraphics[width=180mm]{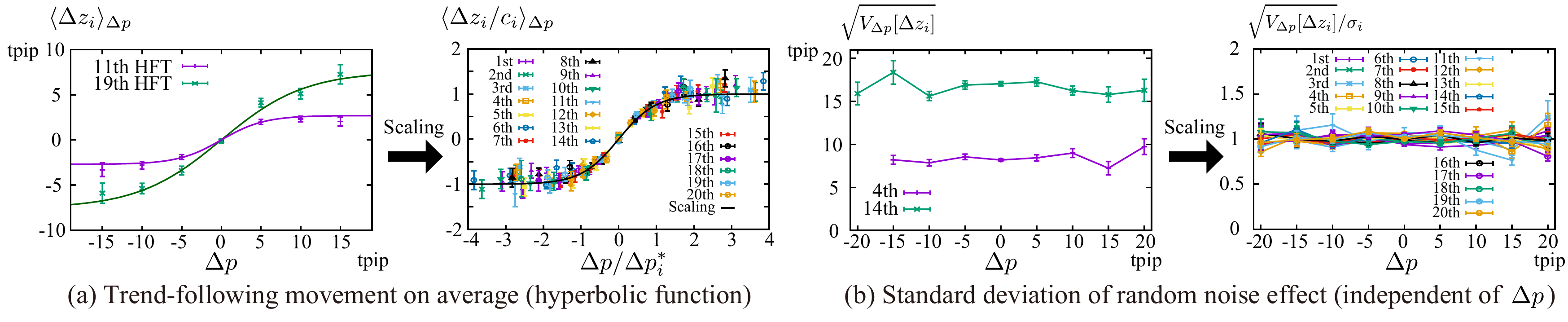}
		\caption	{
							(a)~Average $\Delta z_i$ assuming previous price movements of $\Dp$ and an active trader with $\Delta z_i\neq 0$. 
							The behavior can be fitted by the master curve~{(\ref{eq:trendfollow})} for the top 20 HFTs by introducing scaling parameters $\Dp^*_i$ and $c_i$. 
							(b)~Conditional standard deviation on price movement $\Dp$, showing that the randomness associated with trend following is independent of $\Dp$. 
						}
		\label{fig:trend_follow}
	\end{figure*}
	We then report the empirical microscopic law for the trend-following strategy of individual traders. 
	The bid and ask quoted prices of the top $i$th HFT are denoted by $b_i$ and $a_i$ (see Appendix~\ref{app:trend-following}). 
	We investigated the average movement of the trader's quoted midprice $z_i\equiv(b_i+a_i)/2$ between transactions conditional on 
	the previous market transacted price movement (Fig.~\ref{fig:trajectory_traders}d). 
	Here we introduce the tick time $T$ as an integer time incremented by every transaction.
	The mean transaction interval is $9.3$ seconds during this week. 
	Because typical HFTs frequently modify their price between transactions, we here study HFTs' trend following at one-tick precision. 
	For the top 20 HFTs (Fig.~{\ref{fig:trend_follow}}), we found that the average and variance of movement $\Delta z_i(T)\equiv z_i(T+1)-z_i(T)$ obeyed
	\begin{align}
		\la \Delta z_i \ra_{\Dp} \approx c_i \tanh \frac{\Dp}{\Dp^*_i}, \>\>\>\>\>
		V_{\Dp}[\Delta z_i ]\approx \sigma_i^{2}, \label{eq:trendfollow}
	\end{align}
	where the conditional average $\la \dots \ra_{\Dp}$ is taken when the last price change is $\Dp(T-1)\equiv p(T)-p(T-1)$ and $\Delta z_i\neq 0$ (see Appendix~\ref{app:trend-following}) and 
	the conditional variance is defined by $V_{\Dp}[\Delta z_i ] \equiv \langle \left( \Delta z_i - \langle \Delta z_i\rangle_{\Dp}\right)^2\rangle_{\Dp}$. 
	Here, $p(T)$ is the market transacted price at the $T$ tick,
	and $c_i, \Dp_i^*, \sigma_i^{2}$ are characteristic constants unique to the trader and independent of $\Dp$.
	Their typical values were found to be $c_i\approx 6.0$~tpip, $\Dp_i^* \approx 7.5$~tpip, and $\sigma_i \approx 14.5$~tpip. 
	Our finding~{(\ref{eq:trendfollow})} implies that the reaction of traders is linear for small trends but saturates for large trends,
	and quantifies the collective motion of HFTs. 
	Remarkably, a similar behavior was reported from a price movement data analysis at one-month precision~\cite{Bouchaud2014}.

	{\it Microscopic model.--}
	\begin{figure*}
		\centering
		\includegraphics[width=180mm]{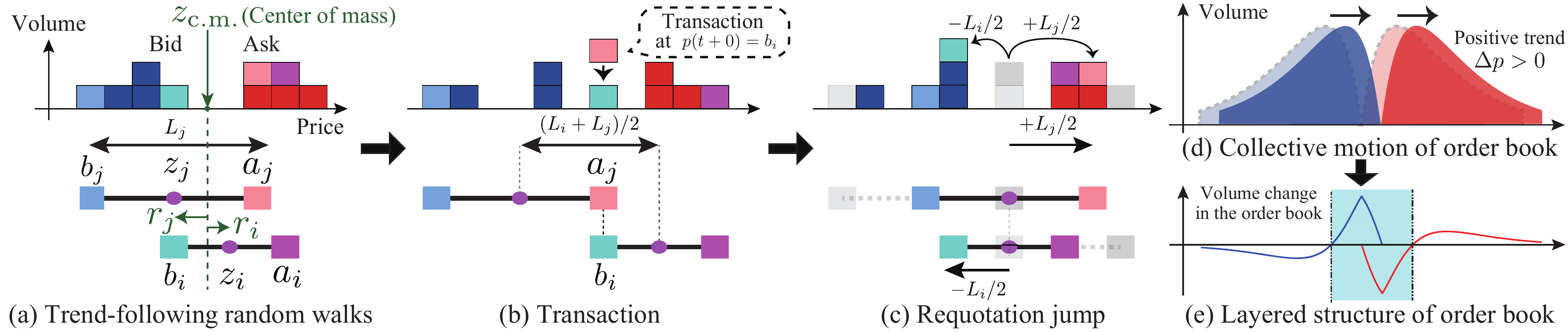}
		\caption{
					Schematic of the microscopic model~{(\ref{eq:dealermodel})}.
					(a)~Midprice of each trader obeys trend-following random walks. 
					(b)~Transaction takes place after price matching $b_i=a_j$ with market transacted price $p$ and its movement $\Dp$ updated. 
					(c)~Pair of traders requote their bid and ask prices simultaneously after transactions. 
					(d)~Order-book collective motion induced by trend following.  
					(e)~Volume change in the bid (ask) order book is positive (negative) near the best price on average for $\Dp>0$.
				}
		\label{fig:DealerModel}
	\end{figure*}
	Here we introduce a minimal microscopic model of HFTs incorporating the above characters. 
	We make four assumptions: (i)~The number of traders is sufficiently large;
	(ii)~traders always quote both bid and ask prices (for the $i$th trader, $b_i$ and $a_i$) simultaneously with a unit volume; 
	(iii)~buy-sell spreads are time constants unique to traders with distribution $\rho(L)$.
	The trader dynamics are then characterized by the midprice $z_i\equiv (b_i+a_i)/2$; and
	(iv)~trend-following random walks are assumed in the microscopic dynamics (Fig.~\ref{fig:DealerModel}a--c),
	\begin{equation}
		\frac{dz_i(t)}{dt} = c\tanh \frac{\Dp(t)}{\Dp^*}
		+ \sigma \eta_i^{\mrR}(t)\label{eq:dealermodel}
	\end{equation}
	with strength for trend following $c$, previous price movement $\Dp$, and white Gaussian noise $\sigma \eta_i^{\mrR}$ with variance $\sigma^{2}$. 
	Here, $c$, $\Dp^*$, and $\sigma$ are assumed shared for all traders for simplicity. 
	In this model, HFTs frequently modify their quoted price by successive submission and cancellation. 
	Indeed, this model can be reformulated as a Poisson price modification process with high cancellation rate (see Appendix~\ref{sec:PoissonPriceModeling}). 
	After transaction $a_j(t)=b_i(t)$ (Fig.~{\ref{fig:DealerModel}b}), 
	the updated market price and its corresponding movement are recorded as 
	\begin{equation}
		p(t+0) = b_i(t), \>\>\>
		\Dp (t+0) = b_i(t) - p(t),
	\end{equation}
	and a requotation jump occurs (Fig.~{\ref{fig:DealerModel}c}),
	\begin{equation}
		z_i(t+0) = z_i(t) - \frac{L_i}{2}, \>\>\> 
		z_j(t+0) = z_j(t) + \frac{L_j}{2}.
	\end{equation} 
	Here, $t+0$ implies the time after transaction. 
	A unique character of this model is the order-book collective motion due to trend following (Fig.~\ref{fig:DealerModel}d). 
	For $\Dp>0$, the bid (ask) volume change tends to be positive (negative) near the best price (Fig.~{\ref{fig:DealerModel}e}),
	consistently with the layered order-book structure~\cite{Yura2014}.
	
	\begin{figure*}
 		\centering
 		\includegraphics[width=180mm]{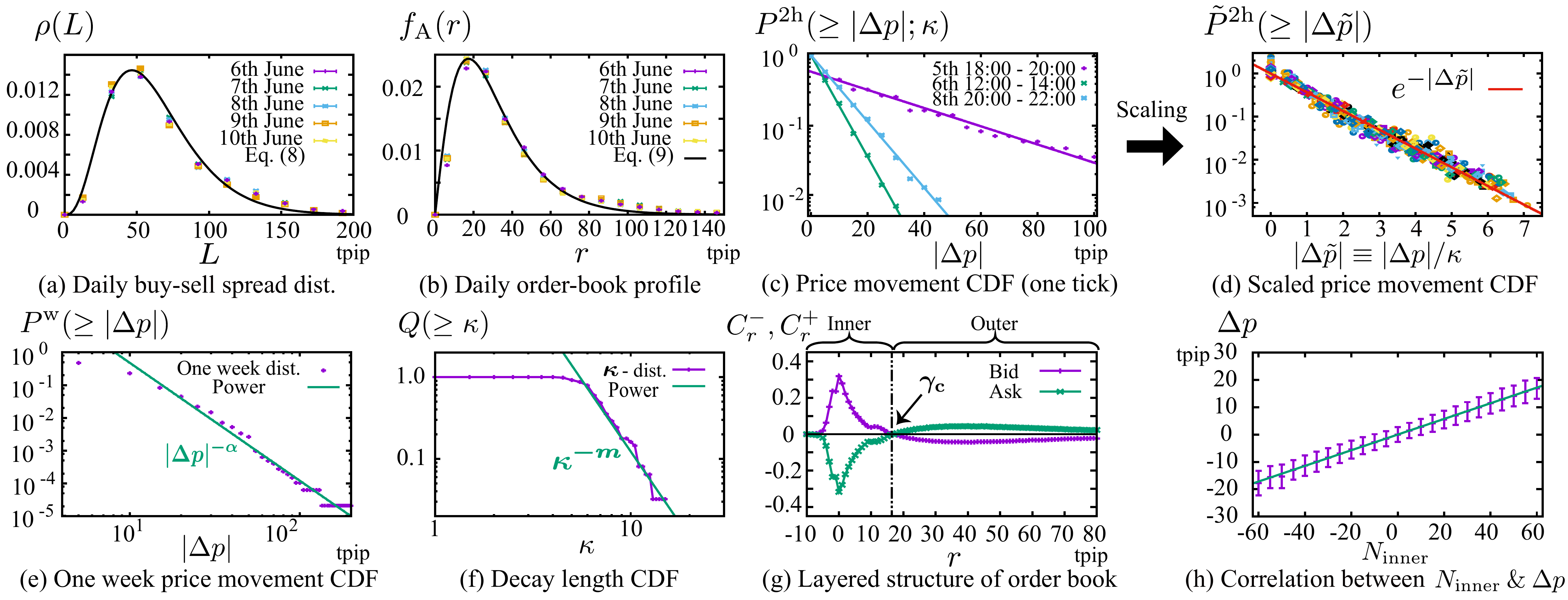}
 		\caption{
					(a)~Daily distribution of buy-sell spreads for HFTs with the empirical master curve~{(\ref{eq:buy-sell})} (see Appendix~\ref{app:buy-sell}). 
 					(b)~Daily average order-book profile for the best prices of HFTs, agreeing with our theoretical line~{(\ref{eq:orderbook})} without fitting parameters (see Appendix~\ref{app:order-book}). 
 					Here the relative depth is measured from the market midprice instead of the c.m. for simplicity, 
					which did not cause big numerical difference in comparing with our theory (see Appendix~\ref{sec:BoltzmannEq}). 
 					(c)~Two-hourly segmented CDFs for the price movement in one-tick precision for the three typical time regions (see Appendix~\ref{app:price_movement}). 
					The CDFs are exponential, consistently with our theoretical prediction~{(\ref{eq:theory_pricemove})}.
					(d)~Two-hourly segmented CDFs are scaled into the single exponential master curve every 2 hour (62 time regions). 
					(e)~Price movement CDF over the whole week obeys a power law of exponent $\alpha$. 
					(f)~Decay length $\kappa$ obeys a power law $Q(\geq \kappa)\sim \kappa^{-m}$.
 					(g)~Order-book layered structure by our HFT model. 
					Pearson's coefficient $C_r^{-} (C_r^+)$ are numerically plotted between $N_r^{-} (N_r^+)$ and $\Dp$ 
					with crossover point $\gamma_{\mathrm{c}}\approx 16.5$~tpip. 
					(h)~Linear correlation between the total number change $N_{\mathrm{inner}}$ in the inner layer and the price movement $\Dp$ with correlation coefficient of $0.63$.
 				}
 		\label{fig:pricediff}
 	\end{figure*}
	
	{\it Kinetic formulation.--}
	We next present an analytical solution to this model~{(\ref{eq:dealermodel})} according to kinetic theory~\cite{Chapman1970,VanKampen}. 
	Let us first introduce the relative distance $r_i\equiv z_i-z_{\CM}$ from the ``center of mass" $z_{\CM}\equiv \sum_{i}z_i/N$ (Fig.~\ref{fig:DealerModel}a),
	where the trend-following effect in Eq.~{(\ref{eq:dealermodel})} is absorbed into the dynamics of $z_{\CM}$.
	The dynamics of $r_i$ become simpler because trend-following effects disappear in this moving frame (see Appendix~\ref{sec:app:CM_RelativePrice}). 
	We next introduce the one-body (two-body) probability distribution as $\phi_L(r)$ ($\phi_{LL'}(r,r')$) conditional on traders' buy-sell spreads.
	From the microscopic model~{(\ref{eq:dealermodel})}, the lowest-order hierarchy equation is derived as
	$\partial \phi_L/\partial t \!=\!(\sigma^{2}/2)(\partial^2 \phi_L/\partial r^2) 
		\!+\! N\!\sum_{s=\pm 1} \!\int \!dL'\rho(L') [J^{s}_{LL'}(r\!+\!sL/2)\!-\!J^{s}_{LL'}]$
	with $J_{LL'}^{s}(r)\!\equiv\! (\sigma^{2}/2)|\tilde{\partial}_{rr'}|\phi_{LL'}(r,r')\big|_{r\!-\!r'\!=s(L\!+\!L')/2}$
	and $|\tilde{\partial}_{rr'}|f\equiv |\partial f/\partial r|+|\partial f/\partial r'|$ (see Appendix~\ref{sec:BBGKY}). 
	By assuming ``molecular chaos" 
	\begin{equation}
		\phi_{LL'}(r,r') \approx \phi_L(r)\phi_{L'}(r'),
	\end{equation}
	we derive the Boltzmann-like equation with collision integrals for the order book: 
	\begin{equation}
		\frac{\partial \phi_L}{\partial t} \!\approx\! \frac{\sigma^{2}}{2}\frac{\partial^2 \phi_L}{\partial r^2} 
		\!+\! N\!\! \sum_{s=\pm 1}\! \int \!\!dL'\rho(L') [\tilde{J}^{s}_{LL'}(r\!+\!sL/2)\!-\!\tilde{J}^{s}_{LL'}]\label{eq:FinancialBoltzmann}
	\end{equation}
	with $\tilde{J}_{LL'}^{s}(r)\!\equiv\!(\sigma^{2}/2)|\tilde{\partial}_{rr'}|\{\phi_{L}(r)\phi_{L'}(r')\}\big|_{r\!-\!r'\!=s(L\!+\!L')/2}$.
	Here, $s=+1$ ($s=-1$) represents transactions as bidder (asker). 
	Because traders exhibit collective motion arising from trend following, a Langevin-like equation is also derived as the macroscopic description of the model~{(\ref{eq:dealermodel})}, 
	\begin{equation}
		\Dp(T+1) = c\tau (T)\tanh \frac{\Dp (T)}{\Dp^*} + \zeta(T), \label{eq:FinancialLangevin}
	\end{equation}
	where $\tau(T)$ and $\zeta(T)$ are transaction time interval and random noise at the $T$th tick time, respectively.
	The first trend-following term corresponds to the momentum inertia in the conventional Langevin equation. 

	Equations~{(\ref{eq:FinancialBoltzmann})} and {(\ref{eq:FinancialLangevin})} can be analytically assessed for $N\to \infty$. 
	We first set the buy-sell spread distribution as 
	\begin{equation}
		\rho(L) = \frac{L^3}{6L^{*4}}e^{-L/L^*}\label{eq:buy-sell}
	\end{equation} 
	with decay length $L^*=15.5\pm 0.2$ tpip, empirically validated in our data set (Fig.~{\ref{fig:pricediff}a}). 
	The solution to Eq.~{(\ref{eq:FinancialBoltzmann})} for $N\to\infty$ is given by 
	$\phi_L(r)=(4/L^2)\max\{L/2-|r|,0\}$. 
	The average order-book profile $f_{\mathrm{A}}(r)=\int dL\rho(L)\phi_L(r-L/2)$ is then given for $r>0$ by
	\begin{equation}
		f_{\mathrm{A}}(r) = \frac{4e^{-\frac{3r}{2L^*}}}{3L^*}\!\!\left[\left(2+\frac{r}{L^*}\right)\sinh \frac{r}{2L^*}\!-\!\frac{re^{-\frac{r}{2L^*}}}{2L^*}\right]. \label{eq:orderbook}
	\end{equation}
	The statistics of $\tau(T)$ in the macroscopic model~{(\ref{eq:FinancialLangevin})} is derived from the mesoscopic model~{(\ref{eq:FinancialBoltzmann})},
	and the tail of the price movement is approximately given by 
	\begin{equation}
		P(\geq |\Dp |;\kappa) \approx e^{-|\Dp |/\kappa} \>\>\>\> (|\Dp |\to \infty)\label{eq:theory_pricemove}
	\end{equation}
	with decay length $\kappa \approx 2\Delta z^*/3$, average movement from trend following $\Delta z^*\equiv c\tau^*$, 
	average transaction interval $\tau^*\approx 3L^{*2}/N\sigma^{2}$, 
	and complementary cumulative distribution function (CDF) $P(\geq |\Dp|;\kappa)$ 
	(see also Appendices~\ref{sec:BoltzmannEq} and \ref{sec:Langevin} for numerical validation). 

	{\it Mesoscopic and macroscopic data analysis.--}
	We next investigated whether our microscopic model is consistent with our data set.
	The empirical daily profile was first studied for the average ask order book for the best prices of HFTs $f_{\mathrm{A}}(r)$ (Fig.~{\ref{fig:pricediff}b}).
 	Surprisingly, we found a quantitative agreement with our theory~{(\ref{eq:orderbook})} without any fitting parameters, which strongly supports the validity of our description. 

	The two-hourly segmented CDF for the price movement is also evaluated in one-tick precision
	$P^{\mathrm{2h}}(\geq |\Dp |;\kappa)$ (Fig.~{\ref{fig:pricediff}c}), 
	which obeys an exponential law that is qualitatively consistent with our theoretical prediction~{(\ref{eq:theory_pricemove})}. 
	The value of the two-hourly decay length $\kappa$ fluctuates significantly during a week. 
	To remove this nonstationary feature, we introduced the two-hourly scaled CDF
	$\tilde{P}^{\mathrm{2h}}(\geq |\Delta \tilde{p}|) \equiv P^{\mathrm{2h}}(\geq \kappa|\Delta \tilde{p}|;\kappa)/Z$ with scaling parameters $\kappa$ and $Z$ (Fig.~{\ref{fig:pricediff}d}),
	thereby incorporating the two-hourly exponential law for the whole week. 

	The price movements obey an exponential law for short periods 
	but simultaneously obey a power law over long periods with exponent $\alpha=3.6\pm0.13$ (Fig.~{\ref{fig:pricediff}e}). 
	This apparent discrepancy originates from the power-law nature of the decay length $\kappa$. 
	Because $\kappa$ approximately obeys a power-law CDF $Q(\geq \kappa)\sim \kappa^{-m}$ over the week with $m=3.5\pm0.13$ (Fig.~\ref{fig:pricediff}f), 
	the one-week CDF $P^{\mathrm{w}}(\geq |\Dp|)$ asymptotically obeys the power law as a superposition of the two-hourly segmented exponential CDF, 
	\begin{equation}
		P^{\mathrm{w}}(\geq |\Dp |) \!=\!\! \int_{0}^\infty \!\!\!d\kappa Q(\kappa)P^{\mathrm{2h}}(\geq |\Dp |;\kappa) \propto |\Dp |^{-m} \label{eq:power-law_weekly}
	\end{equation}
	with $Q(\kappa)\equiv-dQ(\geq \kappa)/d\kappa$, consistently with empirical exponent $\alpha\approx m$. 
	Our result is therefore consistent with the previous reported power law~\cite{Plerou1999,Lux1996,Guillaume1997,Longin1996} as a nonstationary property of $\kappa$. 
	
	Since our trend-following HFT model exhibits the order-book collective motion (Fig.~{\ref{fig:DealerModel}d and e}), 
	this model can reproduce the layered order-book structure~\cite{Yura2014} (see Appendix~\ref{sec:Num_Layers_OB}). 
	Let us define $c_r^-(c_r^+)$ and $a_r^-(a_r^+)$ as the number of bid (ask) submission and cancellation between one tick at the relative distance $r$ from the market midprice. 
	We also define the number change $N_r^{-}=c_r^{-}-a_r^{-}$ $(N_r^{+}=c_r^{+}-a_r^{+})$ at the distance $r$ for the bid (ask) side. 
	Correlation coefficient $C_r^{-}(C_r^{+})$ is plotted in Fig.~{\ref{fig:pricediff}g} between $N_r^{-}(N_r^{+})$ and $\Dp$,
	showing positive and negative correlation in the inner (outer) and outer (inner) layers, respectively. 
	We further show a linear correlation between the price movement $\Dp$ and the total number change in the inner layer $N_{\mathrm{inner}}\equiv\int_{-\infty}^{\gamma_{\mathrm{c}}}dr(N_r^{-}-N_r^{+})$. 
	The trend-following HFT model is thus qualitatively consistent with the previous findings~\cite{Yura2014} (see Appendix~\ref{sec:EmpiricalLayer} for data analyses), 
	implying that the layered structure was the direct consequence of the collective motion. 

	{\it Discussion. --}
	We have empirically studied the trend following of HFTs, inducing the collective motion of the order book. 
	This property has not been captured in the previous order-book model~\cite{Slanina2014,Maslov2000,Daniels2003,Smith2003,Bouchaud2002,Farmer2005} and was critical in reproducing our empirical findings. 
	Indeed, none of our empirical findings, the order-book profile, the exponential price movement, and the layered order-book structure~\cite{Yura2014} were reproduced by the previous order-book model 
	under realistic parameters in the absence of the collective motion (see Appendix~\ref{sec:ZI-OB_Model}). 
	We expect that introduction of this collective motion to order-book models would be the key to replicate these empirical findings.
	
	{\it Conclusion.--}
	We have established both a microscopic model and a kinetic theory for FX traders by direct observation of the HFTs' dynamics, 
	quantitatively agreeing with empirical results under minimal assumptions. 
	In the stream of econophysics, our model~{(\ref{eq:dealermodel})} is the first microscopic model directly supported by microscopic dynamical evidence and exhibiting agreement with mesoscopic and macroscopic findings. 
	We expect that a new stream arises toward systematic descriptions of the financial market based on microscopic evidence. 
	Interested readers are referred to Ref.~\cite{KanazawaFull} for more mathematical details. 

\begin{acknowledgments}
	We greatly appreciate NEX for their provision of the EBS data. 
	We also appreciate M. Katori, H. Hayakawa, S. Ichiki, K. Yamada, S. Ogawa, F. van Wijland, D. Sornette, M. Sano, T.G. Sano, and T. Ito for fruitful discussions. 
	This work was supported by JSPS KAKENHI (Grants No.~16K16016 and No.~17J10781) and JST, Strategic International Collaborative Research Program (SICORP) on the topic of ``ICT for a Resilient Society" by Japan and Israel.
	We thank Richard Haase, Ph.D, from Edanz Group for editing a draft of this manuscript.
\end{acknowledgments}

\appendix

\begin{widetext}
\section{Data Analysis}\label{sec:DataAnalysis}
	
	\subsection{Market rule}\label{sec:MarketRule}
		We analyzed high-frequency trading data in Electronic Broking Services (EBS), one of the biggest financial markets in the world. 
		This market is continuously open except for weekends under few regulations. 
		All trader activities were recorded for our data set with anonymized trader IDs and with one-millisecond time-precision from the 5th 18:00 to the 10th 22:00 GMT June 2016. 
		The minimum price-precision was 0.005 yen for the USD/JPY pair at that time, 
		and the currency unit used in this study is 0.001 yen, called the tenth pip (tpip). 
		The minimum volume unit for transaction was one million USD, and the total monetary flow was about $68$ billion USD during this week. 
		The EBS market is a hybrid market combining both quote-driven and order-driven systems, 
		where traders have three options: limit order, market order, and cancellation. 
		A limit order is an order quoting price with a certain volume and the quoted price displayed on the order book. 
		A market order is an order to buy or sell currencies immediately at the available best price. 
		
		Here we define terminology in this paper.  
		The highest bid and lowest ask quoted prices are called the market best bid and ask prices (denoted by $b_{\mathrm{M}}$ and $a_{\mathrm{M}}$), respectively (see Fig.~\ref{fig:bid_ask_balance}a). 
		The average of the market bid and ask prices is called the market midprice (denoted by $z_{\mathrm{M}}$).  
		Also, the market transacted price $p$ (or the market price for short) means the price at which a transaction occurs in the market.
		
		We note a central trading rule regarding the mutual credit lines between traders~\cite{EBSRule}. 
		All market participants are required to set credit lines to counterparties in advance,
		and they cannot transact with each other in the absence of mutual credit. 
		Therefore, traders sometimes transact at the worse price than the best market price.
		\begin{figure*}
			\centering
			\includegraphics[height = 35mm]{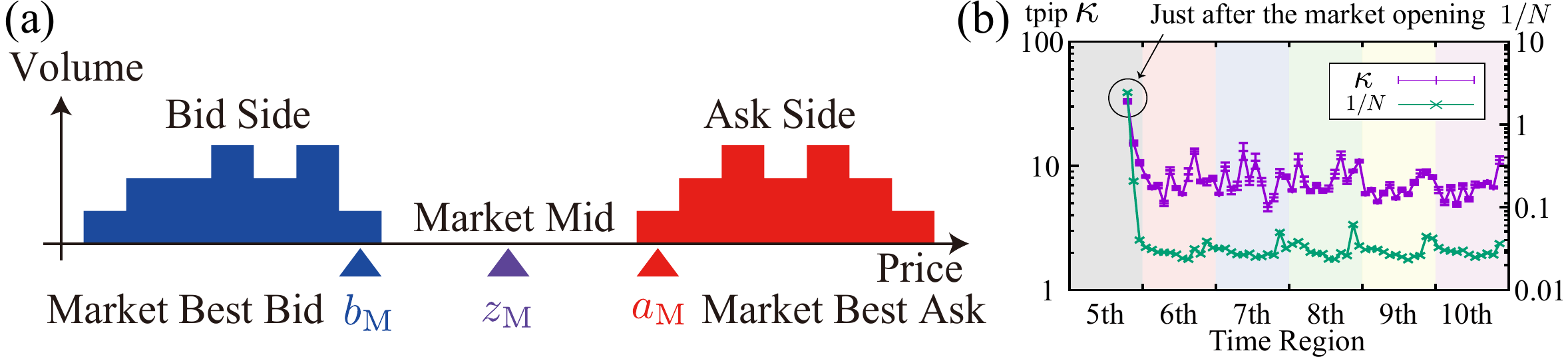}
			\caption	{								
								(a)~Schematic for the market best bid price $b_{\mathrm{M}}$, the market best ask price $a_{\mathrm{M}}$, and the market midprice $z_{\mathrm{M}}$.
								(b)~Time series of the decay length $\kappa$ and the inverse number of HFTs $1/N$ during this week,
								showing that both values of $\kappa$ and $1/N$ were largest during the 5th 18:00 -- 20:00 GMT June (an inactive hour just after the opening of the EBS market).
							}
			\label{fig:bid_ask_balance}
		\end{figure*}

	\subsection{Definition of the high frequency traders}\label{app:def_HFT}
		For this paper, a high frequency trader (HFT) is defined as a trader who submits more than 500 times a day on average (i.e., more than 2500 times for the week). 
		This definition is similar to that introduced in Ref.~\cite{EBS_Schmit}. 
		As a few traders are unwilling to transact and often interrupt orders at the instant of submission, 
		we excluded traders with live orders of less than $0.5\%$ of the transaction time. 
		With this definition, the number of HFTs was 134 during this week, whereas the total number of traders was 1015. 
		We note that the total number of traders who submitted limit orders was 922; the other 93 traders submitted only market orders. 
		We also note that the presence of HFTs has rapidly grown recently and $87.8\%$ of the total orders were submitted by the HFTs in our data set.
		
		Here we note a regulation on cancellations in this market, which is related to motivating HFTs to play the role of key liquidity providers (KLPs)~\cite{EBSRule}. 
		For market stability, all traders are required not to cancel orders frequently; 
		there is a threshold on the ratio between dealt quote and total number of quotations called the quote fill ratio (QFR). 
		If the QFR of a trader is lower than a threshold, penalties are imposed on the trader in this market. 
		However, there is a special rule to lower the threshold. 
		If a trader maintains two-sided quotes continuously for a fixed time interval (called key liquidity hours), the trader qualifies as a KLP and is subject to a lower threshold QFR. 
		Because HFTs tend to cancel orders frequently, they are typically KLPs as illustrated in Fig.~{2a--c}. 
		
		We also note the typical number of HFTs related to snapshots of the order book. 
		We took snapshots of the order book after every transaction and counted the total number of different trader identifiers (IDs) for both bid and ask sides. 
		The counting weight for an HFT quoting both sides is 1 and that for an HFT quoting one side is 1/2. 
		We then plotted the average of the number of trader IDs for both bid and ask sides every two hours in Fig.~{\ref{fig:bid_ask_balance}b},
		showing the periodic intraday activity pattern of HFTs (i.e., $N$ tends to be small during 20:00-22:00 GMT). 
		The typical number of HFT was about $35$ in our data set with this definition. 
		The number of total volumes quoted by HFTs is typically about $80$. 
		Admittedly, there is room for debate on which number is appropriate for the calibration of the total number of traders in our model; 
		it remains a topic for future study.
		
	\subsection{Percentage of two-sided quotes}\label{app:two-sided}
		We calculated the percentage of two-sided quotes as follows; when a bid (ask) order is submitted by a trader, we check whether the corresponding ask (bid) orders exist. 
		We then count the number of two-sided quotes for all traders at the submission of every order and finally divide it by the total number of submissions.

	\subsection{Cancellation ratio for individual traders}\label{app:cancel}
		For each trader, we calculated the total number of canceled volumes over that of submitted volumes for the cancellation ratio of the trader. 
		The cancellation ratio for the first, second, and third top HFTs were 98.59\%, 99.93\%, and 98.70\%, respectively (or equivalently, their QFR were 1.41\%, 0.07\%, and 1.30\%, respectively). 
		The total cancellation ratio among all the HFTs was 94.42\% (or equivalently the total QFR was 5.58\%). 
		
	\subsection{Buy-sell spread}\label{app:buy-sell}
		The difference in the best bid and ask prices was studied as a buy-sell spread for an HFT. 
		Samples where only both bid and ask prices exist are taken at one-second time-intervals for the insets in Fig.~{2a--c} and Fig.~{5a}. 
		We plotted standard deviations of the averages as error bars for each point. 
	
	\subsection{Trend-following effect}\label{app:trend-following}
		We explain the precise definition of the bid (ask) price of individual HFTs for the analysis of trend following. 
		If a trader quotes both single-bid and single-ask orders at any time, the bid and ask prices are defined literally. 
		In the presence of multiple bid or ask orders, we use the best value for the bid or ask orders as $b_i$ or $a_i$. 
		In the absence of any bid or ask or both orders, we use the most recent bid or ask price as $b_i$ or $a_i$ for interpolation. 

		Because of the discrete nature of this data analysis, the probability that traders do not move at all (i.e., $\Delta z_i=0$) is estimated high. 
		We therefore excluded the samples during an inactive time interval $\Delta z_i=0$ for the calculation of representative values in the following. 
		This exception handling does not have a big impact on the hyperbolic structure in Eq.~{(1)}. 	
		Exceptional samples for which the bid or ask price is far from the market price by $0.1$ yen (0.02\% of the total) are also excluded 
		from the calculation of the conditional ensemble average $\la \dots\ra_{\Delta p}$. 
		In Fig.~{3a}, data points are plotted whose samples are over 100 in each bin.
		The standard deviations of the conditional averages are plotted for each point as error bars. 
		Also, median values in the top 20 HFTs are given using $c_i\sim 6.0$~tpip/tick and $\Delta p_i^* \sim 7.5$~tpip, which are estimated by the least squares methods implemented in gnuplot. 

		We have also calculated the standard deviation of quoted price movements for individual traders at one-tick precision in Fig.~{3b}. 
		For the $i$th top HFT, we calculated the conditional variance $V_{\Delta p}[\Delta z_i]\equiv \la (\Delta z_i - \la\Delta z_i\ra_{\Dp})^2\ra_{\Dp}$ and took its square root. 
		As can be seen from Fig.~{3b}, the standard deviation is approximately independent of $\Delta p$ for the top 20 HFTs. 
		We note that the median value was $\sigma_i \sim 14.5$~tpip/tick. 
		This observation is consistent with the assumption that only the drift term depends on $\Dp$ but the random noise effect does not depend on $\Dp$ in our microscopic model. 
		
	\subsection{Average order-book profile}\label{app:order-book}
		The daily average order-book profile is calculated for the best prices of the HFTs. 
		We took snapshots of the order book for the best prices of the HFTs every second and we calculated its ensemble average every day. 
		We also plotted standard deviations of the averages as error bars for each point. 
		
	\subsection{Price movement distributions and decay length}\label{app:price_movement}
		The two-hourly segmented complementary cumulative distribution functions (CDFs) for the price movement $\Delta p$ are calculated in one-tick precision in Figs.~{5c, d}: 
		$\Delta p(T)\equiv p(T+1)-p(T)$ with market price $p(T)$ at tick time $T$. 
		The decay length $\kappa$ and its error were estimated by the least squares methods implemented in gnuplot (Figs.~{5d, f}) and
		the two-hourly scaled CDFs were plotted in Fig.~{5d} with the maximum samples excluded as outliers. 
		The time-series of the estimated decay length $\kappa$ is plotted in Fig.~\ref{fig:bid_ask_balance}b,
		showing that $\kappa$ was the longest just after the opening of the EBS market (the 5th 18:00--20:00 GMT). 
		We conjectured that the decay length $\kappa$ was related to the market activity, 
		represented by such as the number of HFTs during the time region. 
		Indeed, the number of HFTs was also the least during the 5th 18:00--20:00 GMT in the week.

\section{Theoretical Analysis}\label{sec:Theory}

	\subsection{Model dynamics}\label{sec:app:ModelDynamics}
		We explained the model dynamics as trend following random walks~{(2)} with jump rules~{(3)} and~{(4)}. 
		These dynamics can be represented within the framework of Markovian stochastic processes using the $\delta$-functions. 
		The stochastic dynamics can be written as 
		\begin{align}
		\begin{split}
			\frac{dz_i}{dt}		&= c\tanh\frac{\Dp}{\Dp^*} + \sigma \eta_i^{\mrR} + \eta_i^{\mrT}, \>\>\>\>\>\>\>\>\>\>\>\>\>\>\>\>\> 
			\eta_i^{\mrT} \equiv \sum_{k=1}^{\infty}\sum_{j}^{j\neq i}\Delta z_{ij}\delta (t-\tau_{k;ij}),\\
			\frac{dp}{dt} 		&= \sum_{k=1}^\infty \sum_{i,j}^{i<j} (p^{\pst}-p)\delta(t-\tau_{k;ij}),\>\>\> 
			\frac{d\Dp}{dt} 	= \sum_{k=1}^\infty \sum_{i,j}^{i<j} (\Dp^{\pst} -\Dp)\delta(t-\tau_{k;ij}),
		\end{split}
		\label{eq:Model_dynamics}
		\end{align}
		where we have used the It\^o convention. 
		Here, $\tau_{k;ij}$ is the $k$-th collision time; jump size $\Delta z_{ij}$ between traders $i$ and $j$, post-collisional price $p^{\pst}$, and price movement $\Dp^{\pst}$ are defined by
		\begin{equation}
			|z_i(\tau_{k;ij})-z_j(\tau_{k;ij})| = \frac{L_i+L_j}{2} \>\>\> \Longrightarrow
			\Delta z_{ij} = -\frac{L_i}{2}\sgn(z_i-z_j), \>\>\> p^{\pst} = z_i+\Delta z_{ij}, \>\>\> \Dp^{\pst} \equiv z_i+\Delta z_{ij}-p
			\label{eq:CollisonRule}
		\end{equation}
		with signature function $\sgn(x)$ defined by $\sgn(x)=x/|x|$ for $x\neq 0$ and $\sgn(0)=0$. 
		Remarkably, the jump rule Eq.~{(\ref{eq:CollisonRule})} corresponds to the contact condition and momentum exchange in the conventional kinetic theory. 
		In the following, we present effective descriptions of this model for mesoscopic and macroscopic hierarchies. 

	\subsection{Note on a Poisson price modification process}\label{sec:PoissonPriceModeling}
		\begin{figure*}
			\centering
			\includegraphics[height=40mm]{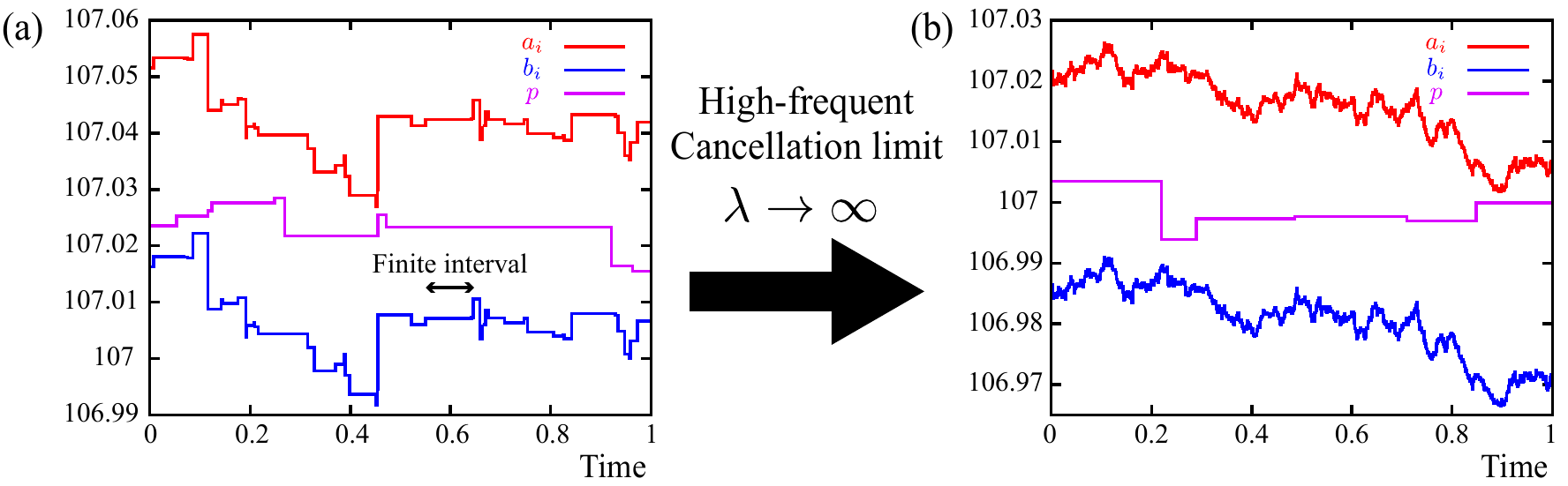}
			\caption	{
								Continuous trend-following random walk model~{(\ref{eq:Model_dynamics})} can be reformulated as the Poisson price modification process~{(\ref{eq:Model_dynamics_discrete})} with high-frequent cancellation rate $\lambda \to \infty$. 
								(a) A typical trajectory of the Poisson price modification process~{(\ref{eq:Model_dynamics_discrete})} with finite cancellation rate $\lambda$. 
								Here the mean interval between price modification by a single trader is set to be $\Dtc\equiv 1/\lambda =\tau^*/4$ with the mean transaction interval $\tau^*$. 
								Other parameters are given by $L^*=15$ tpip, $\Delta p^*=6.75$ tpip, $\Delta z^*=3.6$ tpip, and $N=25$. 
								(b) A typical trajectory of the Poisson price modification process~{(\ref{eq:Model_dynamics_discrete})} with high cancellation rate $\lambda=400/\tau^*$, 
								where the Poisson model~{(\ref{eq:Model_dynamics_discrete})} asymptotically reduces to the continuous model~{(\ref{eq:Model_dynamics})} for $\lambda\to \infty$. 
							}
			\label{fig:schematic_PoissonDealer}
		\end{figure*}
		Since the Gaussian noise can be obtained by taking the high-frequent small jump limit for Poisson noises~\cite{GardinerB},
		the model~{(\ref{eq:Model_dynamics})} can be reformulated as a Poisson price modification process with high-frequent cancellation rate. 
		Here, let us focus on the quoted price dynamics for HFTs in the absence of transactions. 
		As shown in Fig.~{2a--c}, HFTs tend to frequently and continuously modify their price by successive order cancellation and submission,
		possibly due to the market rule (i.e., they are required to maintain the continuous two-sided quote for a fixed time interval~\cite{EBSRule}). 
		On the basis of these characters, we can consider a Poisson cancellation model corresponding to the model~{(\ref{eq:Model_dynamics})}. 
		Let us introduce the order cancellation rate $\lambda$, which gives the cancellation probability during $[t,t+dt]$ as $\lambda dt$. 
		The mean-cancellation interval is characterized by $\Dtc\equiv 1/\lambda$. 
		After cancellation, we assume that HFTs instantaneously requote their price to maintain continuous limit orders. 
		In the absence of transaction, the requoted price is assumed to be described by a discrete version of Eq.~{(\ref{eq:Model_dynamics})} as 
		\begin{equation}
			z_i(t+dt) - z_i(t) =	\begin{cases} 
												0 & (\mbox{Probability = }1-\lambda dt)\\
												c\Dtc \tanh \frac{\Dp}{\Dp^*} + \sigma \sqrt{\Dtc}\eta_i^{\mrR} & (\mbox{Probability = }\lambda dt)
											\end{cases}
			\label{eq:Model_dynamics_discrete}
		\end{equation}
		with a standard Gaussian random number $\eta_i^{\mrR}$ according to our empirical finding~{(1)}. 
		Transaction rule is also assumed the same as the continuous model~{(\ref{eq:Model_dynamics})}. 
		Here, the infinitesimal time step $dt$ is different from the mean-cancellation interval $\Dtc$. 
		A schematic trajectory described by this Poisson dynamics is illustrated in Fig.~\ref{fig:schematic_PoissonDealer}. 
		The continuous model~{(\ref{eq:Model_dynamics})} is obtained in the high-frequent cancellation limit $\lambda \to \infty$ for the discrete model~{(\ref{eq:Model_dynamics_discrete})}. 
		The HFTs' nature on high-frequent price modifications is thus reflected in the continuous model~{(\ref{eq:Model_dynamics})}.

	\subsection{Introduction of the center of mass and the corresponding relative price}\label{sec:app:CM_RelativePrice}	
		\begin{figure}
			\centering
			\includegraphics[height = 40mm]{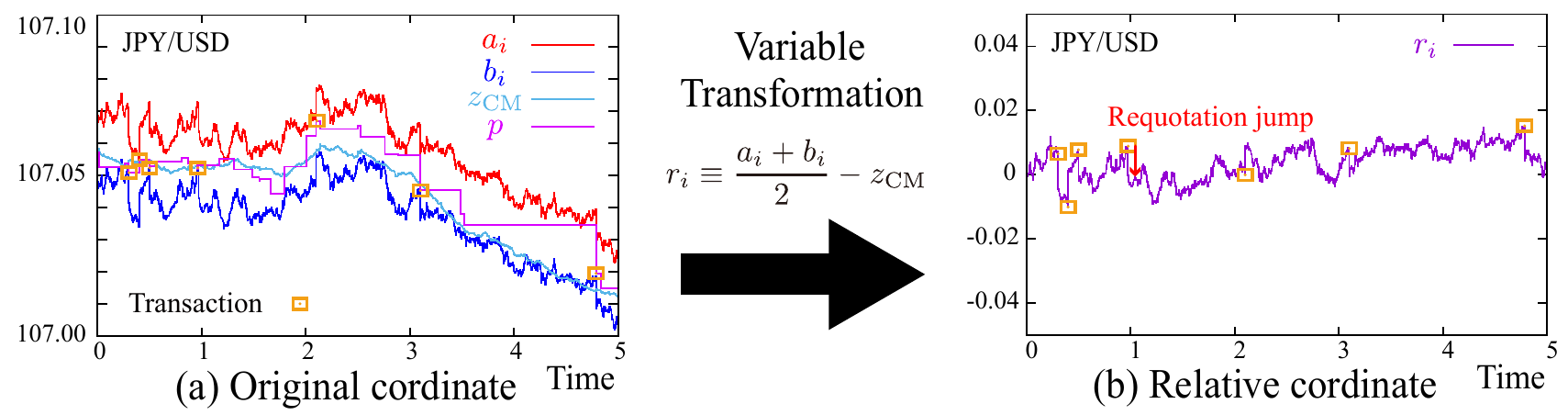}
			\caption	{
									Relative price $r_i\equiv z_i-z_{\CM}$ from the c.m.. 
									(a) The quoted bid and ask prices ($b_i, a_i$) of an individual trader are plotted with the c.m. and market price ($z_{\CM},p$). 
									(b) The trend-following effect is removed in the moving frame $r_i$. 
									The trajectories were obtained from a Monte Carlo simulation of the microscopic model~{(\ref{eq:Model_dynamics})} 
									with time unit $L^{*2}/\sigma^2$, discretized time step $\Delta t = 4.0 \times 10^{-4}L^{*2}/\sigma^{2}$, $L^*=15$ tpip, $\Delta p^*=6.75$ tpip, $\Delta z^*=3.6$ tpip, and $N=25$. 
							}
			\label{fig:RelativePrice}
		\end{figure}
		We here introduce the center of mass (c.m.) and the corresponding relative price (see Fig.~\ref{fig:RelativePrice} for a schematic): 
		\begin{equation}
			z_{\CM} \equiv \frac{1}{N}\sum_{i=1}^N z_i, \>\>\>
			r_i \equiv z_i - z_{\CM}. 
		\end{equation}
		The dynamics of the c.m. and the relative price is given by
		\begin{equation}
			\frac{dz_{\CM}}{dt} = c\tanh\frac{\Dp}{\Dp^*} + \xi, \>\>\> 
			\frac{dr_i}{dt} = \sigma \eta_i^{\mrR} + \eta_i^{\mrT} - \xi, \>\>\>
			\xi\equiv \frac{1}{N} \sum_{j=1}^N \left(\sigma\eta^{\mrR}_j + \eta_j^{\mrT}\right). \label{eq:MacroscopicDynamics}
		\end{equation}
		Remarkably, trend following only appears in the dynamics of the c.m., but does not appear in that of the relative price. 
		This is natural because trend following induces a collective behavior of traders, and can be absorbed into the dynamics of the c.m.. 
		Furthermore, the contribution of $\xi$ is much smaller than that of $\sigma \eta^{\mrR}_i$ and $\eta_{i}^{\mrT}$ for $N\to \infty$: $|\xi| \ll |\sigma \eta_i^{\mrR}+\eta_i^{\mrT}|$. 
		In the moving frame of the c.m., the dynamics of the relative price $r_i$ is thus simplified and approximately obeys the following dynamical equation:
		\begin{equation}
			\frac{dr_i}{dt} \approx \sigma \eta_i^{\mrR} + \eta_i^{\mrT}. 
		\end{equation}	
		
	\subsection{BBGKY Hierarchical equation for two-body problem: $N=2$}\label{sec:two_body_hierarchy}
		Before deriving the Bogoliubov--Born--Green--Kirkwood--Yvon (BBGKY) hierarchical equation for $N\gg 1$, 
		we first consider the two-body system of traders to specify the collision integrals. 
		Extension to the many-body problem will be studied in the next subsection. 
		Let us denote the relative midprices of the first and second traders by $r_1$ and $r_2$ with constant spreads $L_1$ and $L_2$. 
		The dynamics is given by
		\begin{equation}
			\frac{dr_1}{dt} = \sigma\eta_{1;\ve}^{\mrR} + \sum_{k=1}^\infty \Delta r_{1}\delta (t-\tau_{k}), \>\>\>
			\frac{dr_2}{dt} = \sigma\eta_{2;\ve}^{\mrR} + \sum_{k=1}^\infty \Delta r_{2}\delta (t-\tau_{k})
		\end{equation}
		with jump sizes $\Delta r_1$, $\Delta r_2$ and $k$-th transaction time $\tau_k$. 
		Here, $\eta_{i;\ve}$ is the colored Gaussian noise satisfying $\la \eta_{i;\ve}^{\mrR}(t)\eta_{j;\ve}^{\mrR}(s)\ra = \delta_{ij}e^{-|t-s|/\ve}/2\ve$ for $i,j=1,2$. 
		Later, we shall take the $\ve\to 0$ limit, whereby colored Gaussian noise $\eta_{i;\ve}^{\mrR}$ converges to white Gaussian noise as $\lim_{\ve \to 0}\la \eta_{i;\ve}^{\mrR}(t)\eta_{j;\ve}^{\mrR}(s)\ra = \delta_{ij}\delta(t-s)$. 
		The $k$-th transaction time $\tau_k$ and the jump sizes $\Delta r_1$, $\Delta r_2$ are determined using the collision rule,
		\begin{equation}
			|r_1(\tau_k)-r_2(\tau_k)| = \frac{L_1+L_2}{2} \>\>\>
			\Longrightarrow \>\>\>
			\Delta r_1 = -\frac{L_1}{2}\sgn(r_1-r_2), \>\>\>
			\Delta r_2 = -\frac{L_2}{2}\sgn(r_2-r_1).
		\end{equation}
		
		We first derive the master equation for this system. 
		For the two-body probability distribution function (PDF) $P_{12}(r_1,r_2)$, we exactly obtain a time-evolution equation 
		\begin{equation}
			\frac{\partial P_{12}}{\partial t} \!=\! \sum_{i=1,2}\frac{\sigma^2}{2}\frac{\partial^2 P_{12}}{\partial r_i^2} \!+\! 
			\sum_{s=\pm 1} \frac{\sigma^2}{2}\left[\delta(r_1\!-\!r_2)|\tilde{\partial}_{12}|P_{12}\left(r_1+\frac{sL_1}{2},r_2-\frac{sL_2}{2}\right) \!-\! \delta\left(r_1-r_2-s\frac{L_1+L_2}{2}\right)|\tilde{\partial}_{12}|P_{12}\right],
			\label{eq:ME_twobody}
		\end{equation}
		where
		$|\tilde{\partial}_{12}|g(r_1,r_2)\equiv |\partial g(r_1,r_2)/\partial r_1| + |\partial g(r_1,r_2)/\partial r_2|$ is the sum of the absolute value of the partial derivatives for arbitrary $g(r_1,r_2)$. 
		This equation can be derived as follows. 
		For an arbitrary function $f(r_1,r_2)$, we obtain an identity
		\begin{align}
			\frac{df(r_1,r_2)}{dt} &= \sum_{i=1,2}\sigma \eta^{\mrR}_{i;\ve}\frac{\partial f(r_1,r_2)}{\partial r_i} + \sum_{k=1}^\infty\left[ f(r_1+\Delta r_1,r_2+\Delta r_2)-f(r_1,r_2)\right] \delta(t-\tau_k)\notag\\
			&= \sum_{i=1,2}\sigma \eta^{\mrR}_{i;\ve}\frac{\partial f(r_1,r_2)}{\partial r_i} + \sigma\left[ f(r_1+\Delta r_1,r_2+\Delta r_2)-f(r_1,r_2)\right] \delta\left(|r_1-r_2|-\frac{L_1+L_2}{2}\right)|\eta^{\mrR}_{1;\ve}-\eta^{\mrR}_{2;\ve}|,
		\end{align}
		where we have used the expansion of the $\delta$-function: $\delta(g(t)) = \sum_{k=0}^\infty \delta(t-\tau_k)/|g'(\tau_k)|$
		with the $k$-th zero points, such that $g(\tau_k)=0$ and $\tau_k<\tau_{k+1}$. Here we consider the direction of the collision; that is, 
		$\eta_{1;\ve}-\eta_{2;\ve}$ must be positive just before the collision $r_1-r_2=(L_1+L_2)/2$. 
		Inversely, $\eta_{1;\ve}-\eta_{2;\ve}$ must be negative just before the collision $r_1-r_2=-(L_1+L_2)/2$. 
		We thus obtain
		\begin{equation}
			\frac{df}{dt} = \sum_{i=1,2}\sigma \eta^{\mrR}_{i;\ve}\frac{\partial f}{\partial r_i} + \sum_{s=\pm 1}s\sigma\left[ f\left(r_1-\frac{sL_1}{2},r_2+\frac{sL_2}{2}\right)-f\right] \delta\left(r_1-r_2-s\frac{L_1+L_2}{2}\right)(\eta^{\mrR}_{1;\ve}-\eta^{\mrR}_{2;\ve}).
		\end{equation}
		We take the ensemble average of both sides to obtain
		\begin{equation}
			\bigg<\frac{df}{dt}\bigg> = \sum_{i=1,2}\sigma \bigg<\eta^{\mrR}_{i;\ve}\frac{\partial f}{\partial r_i}\bigg> + \sum_{s=\pm 1}s\sigma\bigg<\left[ f\left(r_1-\frac{sL_1}{2},r_2+\frac{sL_2}{2}\right)-f\right] 
			\delta\left(r_1-r_2-s\frac{L_1+L_2}{2}\right)(\eta^{\mrR}_{1;\ve}-\eta^{\mrR}_{2;\ve})\bigg>. 
		\end{equation}
		Here the two-body PDF $P_{12}(x_1,x_2)$ characterizes the probability of $r_1 \in [x_1,x_1+dx_1]$ and $r_2 \in [x_2,x_2+dx_2]$ as $P_{12}(x_1,x_2)dx_1dx_2$. 
		By substituting $f(r_1,r_2)=\delta (r_1-x_1)\delta(r_1-x_2)$, we obtain the master equation 
		\begin{equation}
			\frac{\partial P_{12}}{\partial t} = \sum_{i=1,2}\frac{\sigma^2}{2}\frac{\partial^2 P_{12}}{\partial x_i^2} + 
			\sum_{s=\pm 1} \frac{s\sigma^2}{2}\left[-\delta(x_1-x_2)\tilde{\partial}_{12}P_{12}\left(x_1+\frac{sL_1}{2},x_2-\frac{sL_2}{2}\right)+\delta\left(x_1-x_2-s\frac{L_1+L_2}{2}\right)\tilde{\partial}_{12}P_{12}\right]
		\end{equation}
		where the abbreviation symbol involving the derivatives is defined as $\tilde{\partial}_{12} \equiv \partial/\partial x_1 -\partial/\partial x_2$,
		using the Novikov's theorem~\cite{Novikov} for an arbitrary function $g(r_1,r_2)$ as 
		\begin{equation}
			\lim_{\ve \to 0}\left<\eta^{\mrR}_{i;\ve}(t)g(r_1(t),r_2(t))\right> = \lim_{\ve\to 0}\int_0^t ds\la \eta^{\mrR}_{i;\ve}(t)\eta^{\mrR}_{i;\ve}(s)\ra \left<\frac{\delta g(r_1(t),r_2(t))}{\delta \eta^{\mrR}_{i;\ve}(s)}\right>= \frac{\sigma}{2} \left<\frac{\partial g(r_1,r_2)}{\partial r_i}\right>. 
		\end{equation}
		We note that $\tilde{\partial}_{12}$ is a slightly different symbol from $|\tilde{\partial}_{12}|$ in terms of signatures (see Eq.~{(\ref{eq:relationbtw_absder})} for their relation). 
		We comment on the signature of the derivatives. Considering that $P_{12}(x_1,x_2)\geq 0$ for all $x_1,x_2$ and $P_{12}(x_1,x_2)=0$ for $x_1-x_2>(L_1+L_2)/2$, we obtain 
		$(\partial P_{12}(x_1,x_2)/\partial x_1)|_{x_1-x_2=(L_1+L_2)/2} \leq 0$ and $(\partial P_{12}(x_1,x_2)/\partial x_2)|_{x_1-x_2=(L_1+L_2)/2} \geq 0$. 
		We also obtain $(\partial P_{12}(x_1,x_2)/\partial x_1)|_{x_1-x_2=-(L_1+L_2)/2} \geq 0$ and $(\partial P_{12}(x_1,x_2)/\partial x_2)|_{x_1-x_2=-(L_1+L_2)/2} \leq 0$. 
		In summary, we have
		\begin{equation}
			s\tilde{\partial}_{12}P_{12}(x_1,x_2)\bigg|_{x_1-x_2=s(L_1+L_2)/2} = -|\tilde{\partial}_{12}|P_{12}(x_1,x_2)\bigg|_{x_1-x_2=s(L_1+L_2)/2}.\label{eq:relationbtw_absder}
		\end{equation}
		By a change of notation $x_1\to r_1$ and $x_2\to r_2$, we obtain Eq.~{(\ref{eq:ME_twobody})}. 
		
		By integrating over $r_2$ on both sides, we obtain a hierarchical equation for the one-body PDF $P_1(r_1)\equiv \int dr_2 P_{12}(r_1,r_2)$ as 
		\begin{align}
			\frac{\partial P_{1}(r_1)}{\partial t}
			= \frac{\sigma^2}{2}\frac{\partial^2 P_1(r_1)}{\partial r_1^2} + \sum_{s=\pm 1} \left[J_{12}^s(r_1+sL_1/2) - J_{12}^s(r_1)\right], \>\>\> J_{12}^s(r) \equiv \frac{\sigma^2}{2}|\tilde{\partial}_{12}|P_{12}(r_1,r_2)\bigg|_{r_1-r_2=s(L_1+L_2)/2},
			\label{eq:BBGKY_twobody}
		\end{align}
		where $J_{12}^{s}(r_1)$ is the transaction probability per unit time as bidder $(s=+1)$ or asker $(s=-1)$. 
		The first and second terms on the right-hand side account for the self-diffusion and collision terms, respectively. 
		This is a lowest-order BBGKY hierarchical equation for the special case of $N=2$. 
		Remarkably, the collision term has a quite similar mathematical structure to the collision integral in the conventional Boltzmann equation. 
	
	\subsection{BBGKY hierarchical equation for many-body problem: $N\gg 1$}\label{sec:BBGKY}
		We have derived the hierarchical equation for the one-body PDF for the special case $N=2$. 
		Here we extend the hierarchical equation for the many-body problem with $N \gg 1$. 
		We first assume that the number of traders $N$ is sufficiently large that the spread distribution $\rho(L)$ can be approximated as a continuous function. 
		The one-body and two-body PDFs conditional on buy-sell spread $L$ and $L'$ are denoted by $\phi_L(r)$ and $\phi_{LL'}(r,r')$, respectively. 
		We note the relations $P_i(r_i) = \phi_{L_i}(r_i)$ and $P_{ij}(r_i,r_j) = \phi_{L_iL_j}(r_i,r_j)$ hold for the one-body and two-body PDFs $P_i(r_i)$ and $P_{ij}(r_i,r_j)$ for the traders $i$ and $j$, considering the symmetry between traders. 
		Within the spirit of the Boltzmann equation, 
		the dynamical equation for the one-body distribution $\phi_L(r)$ can be decomposed into two parts: 
		\begin{equation}
			\frac{\partial \phi_L(r)}{\partial t} = \frac{\sigma^{2}}{2}\frac{\partial^2 \phi_L(r)}{\partial r^2}
			+ C(\phi_{LL'}) 
		\end{equation}
		with the self-diffusion term $(\sigma^{2}/2)(\partial^2 \phi_L/\partial r^2)$ and the collision integral $C(\phi_{LL'})$. 
		By extending the collision term in Eq.~{(\ref{eq:BBGKY_twobody})} for large $N \gg1$, 
		we can specify the collision integral as 
		\begin{equation}
			C(\phi_{LL'}) = N\sum_{s=\pm1}\int dL'\rho(L')\left[J^s_{LL'}(r+sL/2)-J^s_{LL'}(r)\right],\>\>\> J^{s}_{LL'}(r) = \frac{\sigma^{*2}}{2}|\tilde{\partial}_{rr'}|\phi_{LL'}(r,r')\bigg|_{r-r'=s(L+L')/2}\label{eq:BBGKY_main}
		\end{equation}
		with the collision probability per unit time as bidder ($s=+1$) or asker ($s=-1$) against a trader with spread $L'$. 
		This is the Boltzmann-like equation, Eq.~{(6)}. 
		We note that this BBGKY hierarchical equation can be systematically derived via the pseudo-Liouville equation.
		The derivation will be given in another technical paper in preparation~\cite{KanazawaFull}.
		
	\subsection{Boltzmann-like equation for finance}\label{sec:BoltzmannEq}
		We next derive a closed equation for the one-body distribution function $\phi_L$ by assuming a mean-field approximation. 
		Let us truncate the two-body correlation (i.e., molecular chaos in kinetic theory),
		\begin{equation}
			\phi_{LL'}(r,r') \approx \phi_L(r)\phi_{L'}(r'). 
		\end{equation}
		A closed mean-field equation for the one-body distribution $\phi_L$ is thereby obtained,
		\begin{equation}
			\frac{\partial \phi_L(r)}{\partial t} = \frac{\sigma^{2}}{2}\frac{\partial^2 \phi_L(r)}{\partial r^2}	+ N\sum_{s=\pm1}\int dL'\rho(L') \left[\tilde{J}^{s}_{LL'}(r+sL/2)-\tilde{J}^{s}_{LL'}(r)\right]\label{eq:BoltzmannEq}
		\end{equation}
		with the mean-field collision probability per unit time as bidder $(s=+1)$ or asker ($s=-1$)
		\begin{equation}
			\tilde{J}^{s}_{LL'}(r) = \frac{\sigma^{2}}{2}|\tilde{\partial}_{rr'}|\left\{\phi_{L}(r)\phi_{L'}(r')\right\} \bigg|_{r-r'=s(L+L')/2}.
		\end{equation}
		Equation~{(\ref{eq:BoltzmannEq})} is a closed equation for the one-body distribution function, and corresponds to the Boltzmann equation in molecular kinetic theory. 
		
		\begin{figure}
			\centering
			\includegraphics[width=160mm]{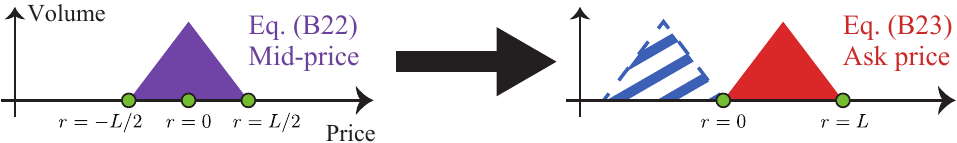}
			\caption	{
								From the trader's midprice order book to the traders' ask order book, with the coordinate shifted by $L/2$. 
							}
			\label{fig:Mid-Ask-Correspondence}
		\end{figure}
		Equation~{(\ref{eq:BoltzmannEq})} can be analytically solved for $N\to \infty$, and the steady solution $\psi_L(r)$ is given by the tent function, 
		\begin{equation}
				\psi_L(r) \equiv \lim_{t\to \infty}\lim_{N\to 0}\phi_L(r;t) = \frac{4}{L^2}\max\left\{ \frac{L}{2}-|r|, 0\right\}.\label{eq:limit_triangular}
		\end{equation}
		Here, a technicality on the appropriate boundary condition will be summarized in another technical paper in preparation~\cite{KanazawaFull}. 
		Note that the tent function~{(\ref{eq:limit_triangular})} for the traders' midprice order book implies the tent functions for both bid and ask order books in shifted coordinates (see Fig.~{\ref{fig:Mid-Ask-Correspondence}} for a schematic).
		The average order-book profile for the ask side $f_{\mathrm{A}}(r)$ is then given by convolution with the tent function,
		\begin{equation}
			f_{\mathrm{A}}(r) = \int dL\rho(L)\psi_L(r-L/2).\label{eq:MF-avg_orderbook}
		\end{equation}
		
		We discuss here the intuitive meaning of the mean-field solution~{(\ref{eq:MF-avg_orderbook})}. 
		The mean-field solution~{(\ref{eq:MF-avg_orderbook})} is exactly zero at $r=\pm L/2$ as $\psi_L(+L/2) = \psi_L(-L/2) = 0$, 
		implying that the edge points $r=\pm L/2$ effectively play the role of hopping barriers at which the particle hops into $r=0$. 
		Indeed, Eq.~{(\ref{eq:limit_triangular})} gives exactly the same solution to the problem of the Brownian motion confined by hopping barriers, as shown in Sec.~\ref{sec_app:IntervalDist}. 
		This is a reasonable result for the $N\to \infty$ limit, where the market is sufficiently liquid and most of the transactions occur just around $r=\pm L/2$. 
		
		\subsubsection{Average order-book profile from the center of mass}
			\begin{figure}
				\centering
				\includegraphics[width =180 mm]{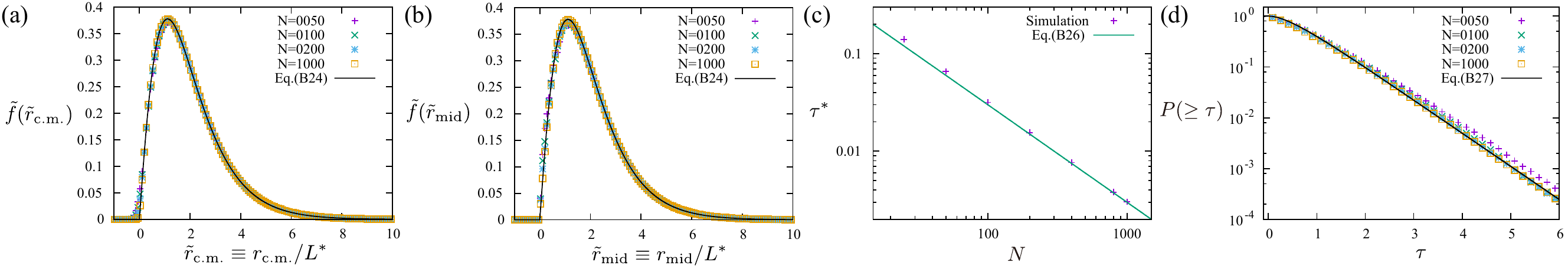}
				\caption	{
								Numerical plots obtained by Monte Carlo simulations of the microscopic model~{(\ref{eq:Model_dynamics})}
								The parameter settings used for the simulation are: $\Delta t = 1.0 \times 10^{-2}L^{*2}/N\sigma^{2}$, $L^*=15$ tpip, $\Delta p^*=4.5$ tpip, $\Delta z^*=3.15$ tpip for various $N$. 
								(a)~Numerical average order-book profiles $f_{\mathrm{A}}^{\CM}(r_{\CM})$ from the c.m. and the theoretical guideline~{(\ref{eq:avg_order_book_2})}. 
								(b)~Numerical average order-book profiles $f_{\mathrm{A}}^{\mathrm{mid}}(r_{\mathrm{mid}})$ from the the market midprice, showing the asymptotic equivalence $f_{\mathrm{A}}^{\CM}(r_{\CM}) \approx f_{\mathrm{A}}^{\mathrm{mid}}(r_{\mathrm{mid}})$. 
								(c)~Numerical mean transaction interval and the theoretical guideline~{(\ref{eq:MeanTransactionInterval})}. 
								(d)~Numerical CDF for transaction interval and the theoretical guideline~{(\ref{eq:transaction_interval})}.
							}
				\label{fig:num_mesoscopic}
			\end{figure}
			If the spreads are distributed in accordance with the $\gamma$-distribution, as empirically studied in the main text, 
			the average order-book profile is given by 
			\begin{equation}
				\rho(L) = \frac{L^{3}e^{-L/L^*}}{6L^{*4}}\>\>\>\Longrightarrow
				f_{\mathrm{A}}(r) = \frac{1}{L^*}\tilde{f}\left(\frac{r}{L^*}\right), \>\>\>
				\tilde{f}(\tr) \equiv \frac{4}{3}e^{-\frac{3\tr}{2}}\left[\left(2+\tr\right)\sinh \frac{\tr}{2}-\frac{\tr}{2}e^{-\frac{\tr}{2}}\right].\label{eq:avg_order_book_2}		
			\end{equation}
			To check the validity of this formula, we performed Monte Carlo simulations of the microscopic model~{(\ref{eq:Model_dynamics})} (Fig.~{\ref{fig:num_mesoscopic}a}),
			where the theoretical formula~{(\ref{eq:avg_order_book_2})} works for various $N$. 
			In the figure, we denote the relative price by $r_{\CM}$ to stress that it is defined from the c.m. as $r_{\CM}\equiv z-z_{\CM}$.  
			
		\subsubsection{Average order-book profile from the market midprice}
			Technically, we have studied the average order-book profile $f_{\mathrm{A}}^{\CM}(r_{\CM})$ from the c.m. instead of that from the market midprice $f_{\mathrm{A}}^{\mathrm{mid}}(r_{\mathrm{mid}})$, 
			because $f_{\mathrm{A}}^{\CM}(r_{\CM})$ is theoretically more tractable than $f_{\mathrm{A}}^{\mathrm{mid}}(r_{\mathrm{mid}})$. 
			Here $r_{\mathrm{mid}}\equiv a_i - z_{\mathrm{M}}$ is the relative distance from the market midprice $z_{\mathrm{M}}$ for the ask price $a_i$ of the $i$th trader. 
			Fortunately, they are asymptotically equivalent for the large $N$ limit and the above formulation is sufficient in understanding the average order-book $f_{\mathrm{A}}^{\mathrm{mid}}(r_{\mathrm{mid}})$ from the market midprice: 
			\begin{equation}
				f_{\mathrm{A}}^{\CM}(r_{\CM}) \approx f_{\mathrm{A}}^{\mathrm{mid}}(r_{\mathrm{mid}}) \>\>\>\>\> (N\to \infty). 
			\end{equation}
			To validate this asymptotic equivalence, we numerically demonstrate the average order-book profile $f_{\mathrm{A}}^{\mathrm{mid}}(r_{\mathrm{mid}})$ from the market midprice $z_{\mathrm{mid}}$ in Fig.~{\ref{fig:num_mesoscopic}b}. 
			This figure numerically shows that the average order-book formula~{(\ref{eq:avg_order_book_2})} is valid even for the order-book from the market midprice.
		
		\subsubsection{Statistics of transaction interval}
			We comment on the statistics of the transaction interval $\tau$. 
			In the mean-field approximation, the average of the transaction interval is given by
			\begin{equation}
				\tau^* \equiv \la \tau \ra \approx \frac{1}{2N\sigma^{2}\int L^{-2}\rho(L) dL} + O(N^{-2}) = \frac{3L^{*2}}{N\sigma^2} + O(N^{-2}),\label{eq:MeanTransactionInterval}
			\end{equation}
			which is phenomenologically derived in Sec.~\ref{sec_app:MeanInterval} and is numerically validated in Fig.~{\ref{fig:num_mesoscopic}c}. 
			Note that this formula can be derived from the pseudo-Liouville equation more systematically~\cite{KanazawaFull}.
			Based on the average transaction interval~{(\ref{eq:MeanTransactionInterval})},	
			the CDF for the  transaction interval $P(\geq \tau)$ is approximately given by the phenomenological formula,
			\begin{equation}
				P(\geq \tau) \equiv \int_{\tau}^\infty d\tau' P(\tau') \approx 1- (1-e^{-3\tau/2\tau^*})^2 \label{eq:transaction_interval}
			\end{equation}
			with transaction interval PDF $P(\tau)$.
			This formula is derived in Sec.~\ref{sec_app:IntervalDist} 
			and is numerically validated in Fig.~{\ref{fig:num_mesoscopic}d}. 
				
	\subsection{Langevin-like equation for finance}\label{sec:Langevin}	
		\begin{figure}
			\centering
			\includegraphics[width =45 mm]{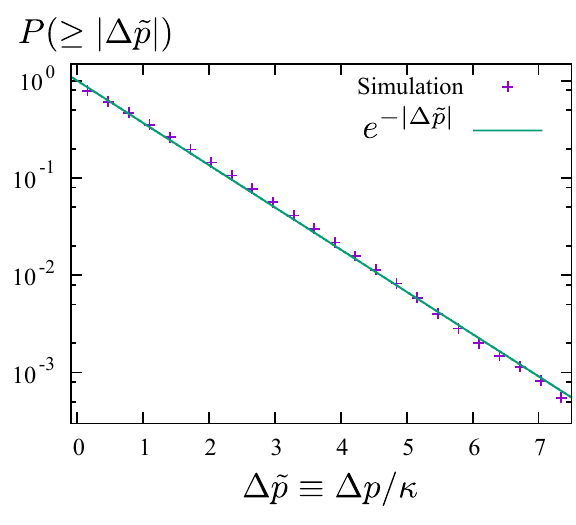}
			\caption	{
								Numerical plots obtained by Monte Carlo simulations of the microscopic model~{(\ref{eq:Model_dynamics})}
								for $\Delta t = 1.0 \times 10^{-4}L^{*2}/\sigma^{2}$, $L^*=15$ tpip, $\Delta p^*=3.0$ tpip, $\Delta z^*=7.2$ tpip for $N=100$. 
							}
			\label{fig:num_pricemovement}
		\end{figure}
		Here we derive phenomenologically the fundamental equation for the financial Brownian motion, which corresponds to the conventional Langevin equation. 
		Let us denote the $T$th transaction price by $p(T)$ and the $T$th price movement by $\Delta p(T)\equiv p(T+1)-p(T)$. 
		Here we focus on the effects of trend following obtained from Eq.~{(\ref{eq:MacroscopicDynamics})}, which induces inertia-like collective motion of the macroscopic dynamics. 
		The dynamical equation for the price movement is thus given by 
		\begin{equation}
			\Delta p(T+1) = c^*\tau(T) \tanh\frac{\Delta p(T)}{\Delta p^*} + \zeta(T),\label{eq:Financial_Brownian_Motion}
		\end{equation} 
		where $\tau(T)$ is the time interval between the $T$th and $(T+1)$th transaction. 
		The first and second terms originate from the trend following and random noise, respectively. 
		Note that the statistics of $\tau(T)$ is derived from the mesoscopic model~{(\ref{eq:BoltzmannEq})} as Eqs.~{(\ref{eq:MeanTransactionInterval})} and~{(\ref{eq:transaction_interval})}. 
		
		Equation~{(\ref{eq:Financial_Brownian_Motion})} governs the macroscopic dynamics of the financial Brownian motion, corresponding to the conventional Langevin equation. 
		For small trend $|\Delta p| \ll \Delta p^*$, we indeed obtain a formal expression similar to the conventional Langevin equation,
		\begin{equation}
			\frac{\Delta^2 p(T)}{\Delta T^2} \approx -\tilde{\gamma}(T)\frac{\Delta p(T)}{\Delta T} + \tilde{\zeta}(T)
		\end{equation}
		with $\tilde{\gamma} (T) \equiv (1-c\tau (T))/\Delta T$, $\tilde {\zeta}(T)\equiv \zeta(T)/(\Delta T)^2$, 
		and $\Delta^2 p(T) \equiv \Delta p(T+1) - \Delta p(T)$.

		We next study the price movement distribution using the financial Langevin equation~{(\ref{eq:Financial_Brownian_Motion})}, which is however a stochastic difference equation that cannot be solved exactly. 
		Nonetheless, its qualitative behavior can be assessed approximately by making the following two assumptions. 
		\begin{enumerate}
			\renewcommand{\labelenumi}{(\roman{enumi})}
			\item The effect of trend following is sufficiently large compared with random noise: $|c^*\tau(T)| \gg |\zeta(T)|$.
			\item The average movement by trend following $\Delta z^*\equiv c\tau^*$ is much larger than the saturation threshold: $\Delta z^* \gg \Delta p^*$.
		\end{enumerate}
		Under condition~{(i)}, the qualitative behavior is governed by the statistics of the transaction interval $\tau(T)$. 
		Under condition~{(ii)}, furthermore, the hyperbolic function in Eq.~{(1)} can be approximated for large fluctuations as $\tanh(\Delta p/\Delta p^*)\approx \sgn(\Delta p)$ and the term $\zeta(T)$ is irrelevant for the tail. 
		Based on Eq.~{(\ref{eq:transaction_interval})} for the transaction interval $\tau$, the price distribution $P(\geq |\Delta p|)$ is approximately obtained for $|\Delta p| \to \infty$ as 
		\begin{equation}
			P(\geq |\Delta p|) \approx e^{-3|\Delta p|/2\Delta z^*}= e^{-|\Delta p|/\kappa},\label{eq:Pc_PriceDiff}
		\end{equation} 
		with an estimated decay length of $\kappa \approx 2\Delta z^*/3$. 
		The validity of this formula was numerically checked in Fig.~{\ref{fig:num_pricemovement}}. 

	\subsection{Numerical analysis of the layered order-book structure for the HFT model}\label{sec:Num_Layers_OB}
		Here we show the detailed analysis to study the layered order-book structure for the HFT model according to the method in Ref.~\cite{Yura2014}. 
		The numerical simulation was performed for the Poisson price modification process~{(\ref{eq:Model_dynamics_discrete})} under the parameter set of 
		$L^*=15.5$ tpip, $\Delta p^*=3.65$ tpip, $\Delta z^*=4.56$ tpip, $\Dtc\equiv 1/\lambda = 4/\tau^*$, and $N=50$. 
		At the instant of an order submission and cancellation for the bid (ask) side, 
		with bin-width of $1$ tpip we measured the relative depth $r$ from the market midprice $z_{\mathrm{M}}$, defined as $r\equiv z_{\mathrm{M}} - b_i$ ($r\equiv a_i - z_{\mathrm{M}}$),
		and we incremented the numbers of $c_r^{-}$ $(c_r^+)$ and $a_r^{-}$ $(a_r^{+})$ by one, respectively. 
		We then accumulated their numbers between $T$ and $T+1$ tick to obtain one sample of $N_r^-(T)\equiv c_r^{-}(T)-a_r^-(T)$ $(N_r^+(T)\equiv  c_r^{+}(T)-a_r^{+}(T))$. 
		We also study the movement of the market price $\Dp(T)\equiv p(T+1)-p(T)$ 
		and calculated Pearson's correlation coefficient $C_r^{-}$ $(C_r^+)$ between $\Dp(T)$ and $N_r^-(T)$ ($N_r^+(T)$) as shown in Fig.~{5g}. 
		The crossover point was estimated to be $\gamma_{\mathrm{c}}\approx 16.5$ tpip in the numerical simulation. 

		We next study the linear correlation between the number change $N_{\mathrm{inner}}$ in the inner layer and the price movement $\Dp$. 
		Between $T$ and $T+1$ tick, we take a sample of both $N_r^-(T)$ and $N_r^+(T)$, and calculated their integral in the inner layer as $N_{\mathrm{inner}}(T)\equiv \int_{-\infty}^{\gamma_{\mathrm{c}}}dr \{N_r^{-}(T)-N_r^{+}(T)\}$. 
		We then calculated the correlation between $N_{\mathrm{inner}}(T)$ and $\Dp(T)$ for Fig.~{5h}. 
		We have plotted the average of $\Dp(T)$ conditional on $N_{\mathrm{inner}}(T)$ with errorbars representing the conditional variance. 
		The figure shows their significant linear correlation of Pearson's coefficient $0.63$. 
	
	\subsection{Empirical analysis of the layered structure of the order book in the data set}\label{sec:EmpiricalLayer}
		\begin{figure}
			\centering
			\includegraphics[height=40mm]{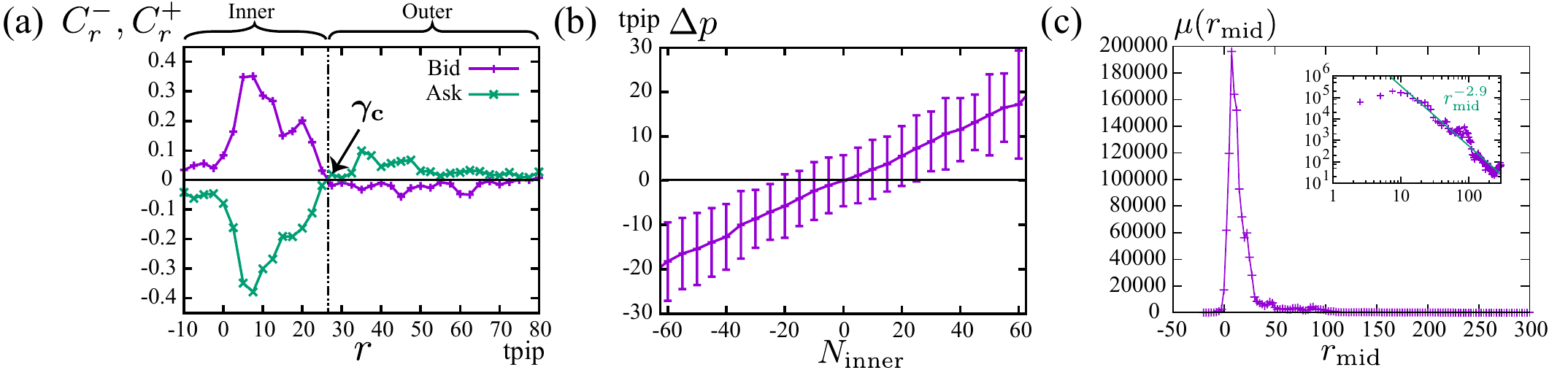}
			\caption	{
							(a) The empirical layered structure of the HFT's order book in the data set, showing the crossover point $\gamma_{\mathrm{c}}\approx 26$ tpip between the inner and outer layers.
							(b) The linear correlation between the total number change in the inner layer $N_{\mathrm{inner}}$ and the market transacted price movement $\Dp$. 
							Pearson's correlation coefficient is given by $0.616$ between $N_{\mathrm{inner}}$ and $\Dp$. 
							(c) The distribution of limit order submissions from the market midprice, showing a power-law tail with exponent $2.9$ (see the inset for the log-log plot). 
						}
			\label{fig:empirical_layered_structure}
		\end{figure}
		We show the empirical layered structure of the HFTs' order book in our data set. 
		According to the essentially same method in Sec.~\ref{sec:Num_Layers_OB}, we have calculated the layered structure as shown in Fig.~\ref{fig:empirical_layered_structure}a and b. 
		The volume change in the inner layer $N_{\mathrm{inner}}(T)$ has a significant correlation of Pearson's coefficient $0.616$ with the price movement $\Dp(T)$.
		
		For consistency throughout this Letter, we have focused on the best prices of HFTs for the correlation analysis in Fig.~\ref{fig:empirical_layered_structure}a and b. 
		In other words, we incremented $c_r^{-}$ and $c_r^{+}$ when the newly quoted price was the best price of the trader. 
		Also, we incremented $a_r^{-}$ and $a_r^{+}$ when the price of the canceled order was the best price of the trader. 
	
	\subsection{Numerical comparison with the zero-intelligence order-book models}\label{sec:ZI-OB_Model}
		\begin{figure}
			\centering
			\includegraphics[width=170mm]{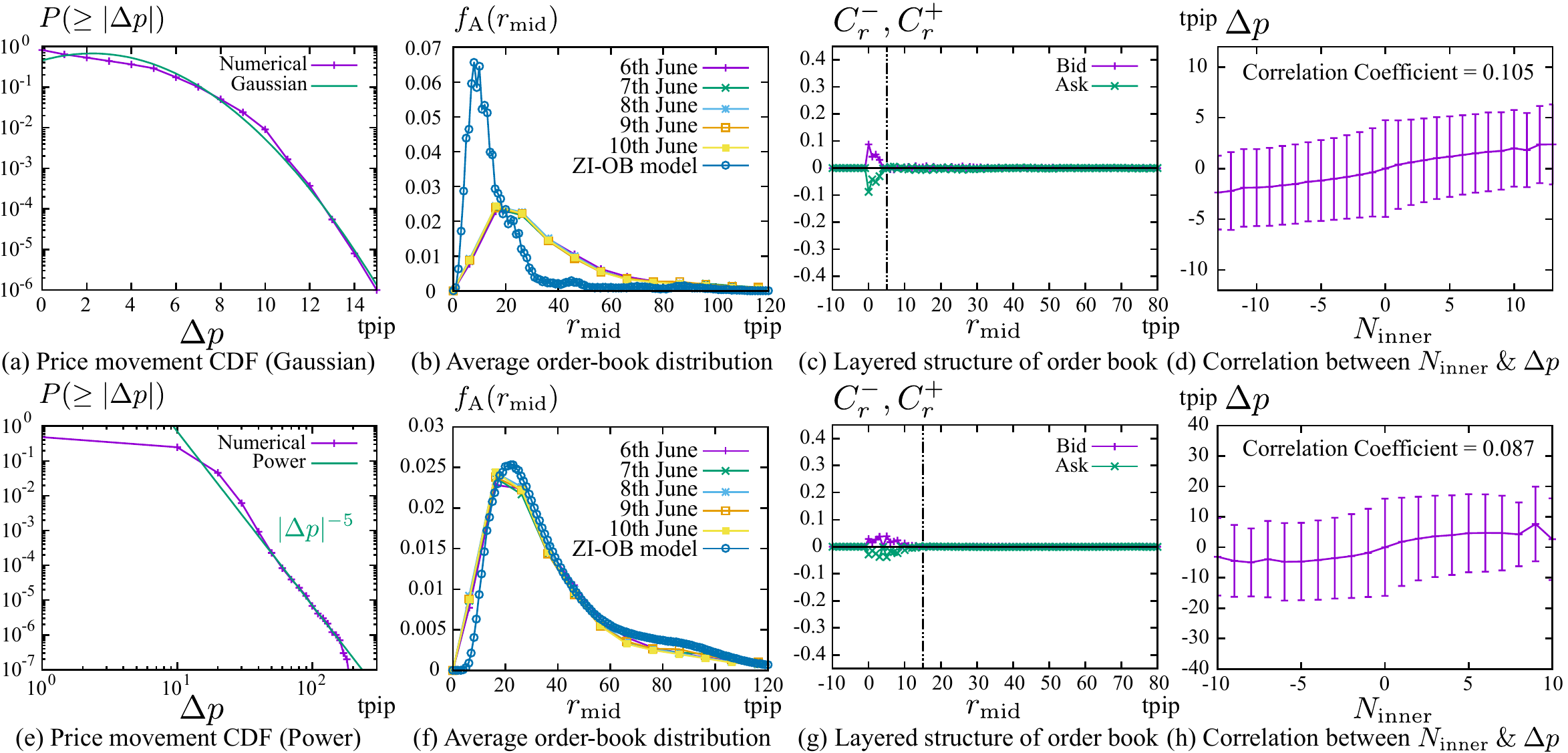}
			\caption{
						Numerical study on the ZI-OB model~\cite{Daniels2003,Smith2003,Bouchaud2002} to examine its consistency with the empirical findings in the main text. 
						(a--d) We first studied a simulation under a realistic parameter set satisfying $(N_{\mathrm{vol}}, \mathrm{QFR}) \approx (100, 5\%)$. 
						The price movement obeyed the Gaussian law (Fig.~a), which is contradictory to the exponential law in our data set. 
						The average order-book profile did not quantitatively fit the real order book during this week (Fig.~b). 
						The layered order-book structure was not also observed (Figs.~c and d). 
						(e--h) By adjusting the market order rate $\omega$, we attempted to fit the real order-book profile by the ZI-OB model. 
						Though the average order-book profile was replicated by the ZI-OB model by parameter adjustment (Fig.~f), 
						neither the price movement statistics nor the layered order-book structure were consistent with our data set instead (i.e., Fig.~e shows the power-law statistics for $\Dp$ and Figs.~g and h shows the absence of the layered structure). 
						We also note that the adjusted parameter implies $\mathrm{QFR}\approx 75\%$, which is over ten-times larger than the real QFR. 
						In this sense, the ZI-OB model was not consistent with the empirical findings under realistic parameters. 
			}
			\label{fig:zero_intelligence}
		\end{figure}
		Here we compare our empirical findings with the zero-intelligence order-book (ZI-OB) model~\cite{Daniels2003,Smith2003,Bouchaud2002}. 
		The basic ZI-OB model is the uniform decomposition order-book model introduced in Ref.~\cite{Daniels2003,Smith2003},
		where both submission and cancellation are assumed to obey the homogeneous Poisson processes. 
		To understand real average order-book profiles, in Ref.~\cite{Bouchaud2002}, the uniform submission rate was replaced with a real nonuniform submission rate obeying a power law. 
		Here we study the improved ZI-OB model in Ref.~\cite{Bouchaud2002} from the viewpoint of the consistency with our empirical findings.
		The inputs to the ZI-OB model are the following three components. 
		\begin{enumerate}
			\item Submission rate density $\mu(r_{\mathrm{mid}})$: limit order submissions are assumed to obey the inhomogeneous Poisson process characterized by the submission rate $\mu(\rmid)$ with relative depth $\rmid$ from the market midprice.
					In other words, a new limit order is submitted in the range $[\rmid,\rmid+d\rmid]$ between time interval $[t,t+dt]$ with probability of 
					\begin{equation}
						\mu(r_{\mathrm{mid}})d\rmid dt.
					\end{equation}
					The empirical submission histogram in our data set is depicted in Fig.~\ref{fig:empirical_layered_structure}c, showing a power-law tail with exponent $2.9$. 
					For our numerical implementation, the limit order submission rate is directly fixed from the empirical submission histogram for $\rmid\geq 0$. 
					The total submission rate is given by 
					\begin{equation}
						\mutot \equiv \int_0^\infty d\rmid \mu(\rmid),
					\end{equation}
					characterizing the frequency of total submissions. 
					The gender of order (i.e., buy or sell) is randomly selected with equal probability. 
			\item Cancellation rate $\lambda$: any order is assumed to be cancel according to the homogeneous Poisson process with intensity $\lambda$. 
					In other words, an order is canceled between time interval $[t,t+dt]$ with probability of 
					\begin{equation}
						\lambda dt. 
					\end{equation}
			\item Market order rate $\omega$: market orders are assumed to obey the Poisson process with intensity $\omega$. 
					In other words, a buy or sell market order is submitted between interval $[t,t+dt]$ with probability of 
					\begin{equation}
						\omega dt. 
					\end{equation}
					The gender of order is randomly selected with equal probability. 
		\end{enumerate}
		These parameters characterize the order-book dynamics in the steady state. 
		For example, the average total order-book volume $N_{\mathrm{vol}}$ in both sides and the QFR (i.e., the probability for an order to be transacted finally) are given by
		\begin{equation}
			N_{\mathrm{vol}} \approx \frac{\mutot - \omega}{\lambda}, \>\>\> \mathrm{QFR} \approx \frac{\omega}{\mutot},
		\end{equation}
		respectively. 
		These relations are deduced from the conservation of order flux in the steady state: $\mutot \approx \lambda N_{\mathrm{vol}} + \omega$. 

		\subsubsection{Numerical simulation with realistic parameters}
			We first consider a numerical simulation based on realistic parameters in our data set with minimum price precision of $1$ tpip.  
			The submission rate density $\mu(\rmid)$ is directly fixed from the empirical submission histogram in our data set (Fig.~\ref{fig:empirical_layered_structure}c). 
			The cancellation and market order rates are fixed as $\lambda/\mutot = 9.5\times 10^{-3}$ and $\omega/\mutot = 0.05$ 
			to satisfy $N_{\mathrm{vol}}\approx 100$ and $\mathrm{QFR}\approx 5\%$. 
			Though these parameters were realistic in our data set, the numerical results in Fig.~\ref{fig:zero_intelligence}a--d were not consistent with the empirical findings.
			Indeed, the price movement in the ZI-OB model obeyed the Gaussian statistics (Fig.~\ref{fig:zero_intelligence}a), which is different from the empirical exponential law in our data set. 
			The numerical average order-book profile did not quantitatively fit the real order-book profile (Fig.~\ref{fig:zero_intelligence}b). 
			In addition, the layered structure of the order book did not emerge from the ZI-OB model (Fig.~\ref{fig:zero_intelligence}c and d). 
			To replicate these empirical findings, in particular the layered order-book structure, 
			we conjectured that the collective motion of the limit order book (i.e., the microscopic trend-following behavior) needs to be incorporated with conventional order-book models.  

		\subsubsection{Numerical simulation with adjusted parameters}
			In Ref.~\cite{Bouchaud2002}, the possibility of the ZI-OB model was studied to fit realistic order-book profiles by adjusting parameters. 
			In the same way, we here seek the possibility to adjust the model parameters in replicating the real order-book profile in our data set. 
			By fixing the average order-book volume as $N_{\mathrm{vol}}\approx 100$, 
			we adjusted the market order rate $\omega$ as a fitting parameter to replicate the real order-book profile in our data set (see Fig.~\ref{fig:zero_intelligence}e--h). 
			By inputting $\omega/\mutot=0.75$ (or equivalently $\mathrm{QFR}\approx 75\%$), the ZI-OB model replicated the real order-book profile as shown in Fig.~\ref{fig:zero_intelligence}f. 
			Instead, however, other numerical results of the ZI-OB model were not consistent with the exponential price movement statistics (Fig.~\ref{fig:zero_intelligence}e for the power-law price movement) nor the layered order-book structure (Fig.~\ref{fig:zero_intelligence}g and h).
			In addition, the parameter adjustment implied $\mathrm{QFR}\approx 75\%$, which was over ten-times larger than the real QFR (see Sec.~\ref{app:cancel}). 
			We thus conclude that our empirical results were not consistently replicated by the ZI-OB model under realistic parameters, at least in our data set. 

\section{Technical Issues for Derivation}\label{sec:Technical}

	\subsection{Brownian motion confined by hopping barriers}\label{sec_app:MeanInterval}
		In this subsection, we study the Brownian motion confined by the hopping barriers at $r=\pm L/2$ (see Fig.~\ref{fig:BrownianWithHopBarriers} for a schematic). 
		Let us assume that a particle moves randomly in the absence of collision for $r \in (-L/2,L/2)$. 
		We then place hopping barriers at $r=\pm L/2$, and we assume that the particle moves to the origin $r=0$ after collisions.
		The particle's position $r(t)$ then obeys the dynamical equation
		\begin{equation} 
			\frac{dr}{dt} = \sigma \heta^{\mrR} + \heta^{\mrT}_+ + \heta^{\mrT}_-, \>\>\>
			\heta^{\mrT}_+ = -\frac{L}{2}\sum_{k=1}^\infty \delta (t-\htau_{k}^+), \>\>\>
			\heta^{\mrT}_- = +\frac{L}{2}\sum_{k=1}^\infty \delta (t-\htau_{k}^-),\label{eq:BMbwHopBarriers}
		\end{equation}
		where $\heta^{\mrR}$ is the white Gaussian noise with unit variance, and 
		$\heta^{\mrT}_+$ and $\heta^{\mrT}_-$ are respectively the jump terms originating from the hopping barriers at $r=\pm L/2$. 
		Here, the $k$th collision times $\tau_{k}^+$ and $\tau_{k}^-$ at the barriers $r=\pm L/2$ satisfy the relation $\hr(\tau_k^{\pm})=\pm L/2$. 
		In a parallel calculation to that in Sec.~\ref{sec:two_body_hierarchy},
		the dynamical equation for the probability distribution function $P(r)$ is given by
		\begin{equation}
			\frac{\partial P(r)}{\partial t} = \frac{\sigma^{2}}{2}\frac{\partial^2 }{\partial r^2}P(r) + \sum_{s=\pm 1} [J^s(r-sL/2)-J^s(r)], \>\>\>
			J^s (r) = \frac{\sigma^{2}}{2}\delta(r+sL/2)|\partial_{s}P(r)|.
		\end{equation}
		The steady solution is then given by the tent function, $P_{\mathrm{SS}}(r) \equiv \lim_{t\to \infty}P(r) = \psi_L(r),$
		which is the same as the mean-field solution~{(\ref{eq:limit_triangular})} for $N\to \infty$. 
		This implies that the mean-field description corresponds to Brownian motion confined by the hopping barriers in the limit $N\to \infty$. 
		\begin{figure}
			\centering
			\includegraphics[width=75mm]{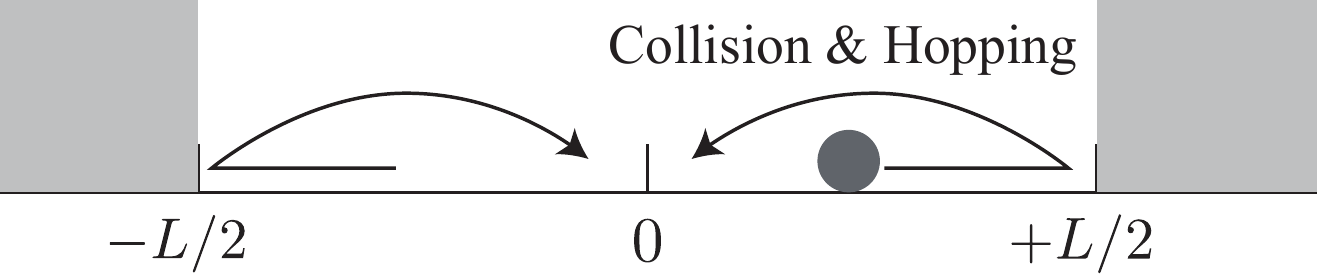}
			\caption{		
						Schematic of Brownian motion confined by the hopping barriers at $r=\pm L/2$. 
						When the Brownian particle collides with the hopping barriers, the particle hops to the origin $r=0$. 
					}
			\label{fig:BrownianWithHopBarriers}
		\end{figure}

		From the above picture, we can derive Eq.~{(\ref{eq:MeanTransactionInterval})} for the mean transaction interval asymptotically in terms of $N$. 
		Because a single particle in our model behaves as the Brownian motion confined by the hopping barriers for $N\to \infty$, 
		the mean transaction interval for a single particle can be derived by considering the survival rate problem for the model~{(\ref{eq:BMbwHopBarriers})}. 
		According to Ref.~\cite{Yamada2009}, the mean transaction interval is given by $L^2/4\sigma^{2}$ for a single particle. 
		We next derive the mean transaction interval for the whole system. 
		A count of the number of collisions $n_L$ for the spread $L$ during the time interval $T$ yields $n_L = T/(L^2/4\sigma^{2})$, 
		when $T$ is sufficiently large. 
		The total number of collisions $n_{\mathrm{tot}}$ is then given by
		\begin{equation}
			n_{\mathrm{tot}} = \sum_{i=1}^N \frac{T}{L_i^2/4\sigma^{2}} \approx N\int \frac{dL\rho(L) T}{L^2/4\sigma^{2}},
		\end{equation}
		where there are duplicate counts because any transaction occurs as a binary collision. 
		Considering the duplicate counts, the mean transaction interval $\tau^*$ for the whole system is given by
		$\tau^* = T/(n_{\mathrm{tot}}/2)$, which implies Eq.~{(\ref{eq:MeanTransactionInterval})}.
	
\subsection{Transaction interval distribution}\label{sec_app:IntervalDist}
	The phenomenological estimation of the cumulative distribution for transaction interval~{(\ref{eq:transaction_interval})} is presented here. 
	Let us assume that the arrival-time intervals of a bidder and an asker at the center of mass obey the Poisson statistics:
	\begin{equation}
		P_{\mathrm{A}}(\geq \tau_{\mathrm{A}}) = \int_{\tau_{\mathrm{A}}}^\infty P_{\mathrm{A}}(\tau'_{\mathrm{A}}) d\tau'_{\mathrm{A}} = e^{-\tau_{\mathrm{A}}/a},\>\>\>
		P_{\mathrm{B}}(\geq \tau_{\mathrm{B}}) = \int_{\tau_{\mathrm{B}}}^\infty P_{\mathrm{B}}(\tau'_{\mathrm{B}}) d\tau'_{\mathrm{B}} = e^{-\tau_{\mathrm{B}}/a}
	\end{equation}
	with the characteristic time interval $a$. 
	$P_{\mathrm{A}}(\tau_{\mathrm{A}})$ ($P_{\mathrm{B}}(\tau_{\mathrm{B}})$) and $P_{\mathrm{A}}(\geq \tau_{\mathrm{A}})$ ($P_{\mathrm{B}}(\geq \tau_{\mathrm{B}})$) are the PDFs and CDFs of arrival time intervals 
	for an asker (a bidder), respectively.
	We also assume that the transaction occurs when both bidder and asker arrive at the center of mass. 
	This picture implies that the transaction interval $\tau$ is approximately given by
	\begin{equation}
		\tau \approx \max \{\tau_{\mathrm{A}}, \tau_{\mathrm{B}}\} 
		\Longrightarrow 
		P(\geq \tau) = 1-(1-e^{-\tau/a})^2, \label{app:eq:Max_Interval}
	\end{equation}
	where we have used a formula for the order statistics~\cite{David2003}. 
	Considering the consistency between Eq.~{(\ref{app:eq:Max_Interval})} and the mean transaction interval~{(\ref{eq:MeanTransactionInterval})}, 
	we obtain the self-consistent condition $\tau^*=3a/2$. 
	Equation~{(\ref{eq:transaction_interval})} then follows. 

\end{widetext}

\end{document}